%% file: CEMP_clusters.tex
\documentclass[twocolumn]{aastex631}
\usepackage{amsmath}
\usepackage{mathtools}
\usepackage{multirow}
\usepackage{longtable}
\usepackage{afterpage}
\usepackage{amssymb}

\usepackage{array}

\shorttitle{Chemo-Dynamically Tagged Groups of Halo CEMP Stars}

\begin{document}

\title{Chemo-Dynamically Tagged Groups of CEMP Stars in the Halo of the Milky Way. I.\\ Untangling the Origins of CEMP-$s$ and CEMP-no Stars}

\author[0000-0003-0998-2744]{Joseph Zepeda}
\affiliation{Department of Physics and Astronomy, University of Notre Dame, Notre Dame, IN 46556, USA}
\affiliation{Joint Institute for Nuclear Astrophysics -- Center for the Evolution of the Elements (JINA-CEE), USA}

\author[0000-0003-4573-6233]{Timothy C. Beers}
\affiliation{Department of Physics and Astronomy, University of Notre Dame, Notre Dame, IN 46556, USA}
\affiliation{Joint Institute for Nuclear Astrophysics -- Center for the Evolution of the Elements (JINA-CEE), USA}

\author[0000-0003-4479-1265]{Vinicius M. Placco}
\affiliation{NSF’s NOIRLab, 950 N. Cherry Ave., Tucson, AZ 85719, USA}

\author[0000-0001-9723-6121]{Derek Shank}
\affiliation{Department of Physics and Astronomy, University of Notre Dame, Notre Dame, IN 46556, USA}
\affiliation{Joint Institute for Nuclear Astrophysics -- Center for the Evolution of the Elements (JINA-CEE), USA}

\author[0000-0003-3246-0290]{Dmitrii Gudin}
\affiliation{Department of Mathematics, University of Maryland, College Park, MD 20742-4015 USA}

\author[0000-0002-5661-033X]{Yutaka Hirai}
\altaffiliation{JSPS Research Fellow}
\affiliation{Department of Physics and Astronomy, University of Notre Dame, Notre Dame, IN 46556, USA}
\affiliation{Astronomical Institute, Tohoku University, 6-3 Aoba, Aramaki, Aoba-ku, Sendai, Miyagi 980-8578, Japan}
\affiliation{Joint Institute for Nuclear Astrophysics -- Center for the Evolution of the Elements (JINA-CEE), USA}

\author{Mohammad Mardini}
\affiliation{Kavli IPMU (WPI), UTIAS, The University of Tokyo, Kashiwa, Chiba 277-8583, Japan}
\affiliation{Institute for AI and Beyond, The University of Tokyo 7-3-1 Hongo, Bunkyo-ku, Tokyo 113-8655, Japan}

\author{Colin Pifer}
\affiliation{Department of Physics and Astronomy, University of Notre Dame, Notre Dame, IN 46556, USA}
\affiliation{Joint Institute for Nuclear Astrophysics -- Center for the Evolution of the Elements (JINA-CEE), USA}

\author{Thomas Catapano}
\affiliation{Department of Physics and Astronomy, University of Notre Dame, Notre Dame, IN 46556, USA}
\affiliation{Joint Institute for Nuclear Astrophysics -- Center for the Evolution of the Elements (JINA-CEE), USA}

\author{Sean Calagna}
\affiliation{Department of Physics and Astronomy, University of Notre Dame, Notre Dame, IN 46556, USA}
\affiliation{Joint Institute for Nuclear Astrophysics -- Center for the Evolution of the Elements (JINA-CEE), USA}




\shortauthors{Zepeda et al.}

\begin{abstract}

We construct a sample of 644 carbon-enhanced metal-poor (CEMP) stars with abundance analyses based on moderate- to high-resolution spectroscopic studies. Dynamical parameters for these stars are estimated, based on radial velocities, Bayesian parallax-based distance estimates, and proper motions from $Gaia$ EDR3 and DR3, supplemented by additional available information where needed. After separating our sample into the different CEMP morphological groups in the Yoon-Beers Diagram of absolute carbon abundance vs. metallicity, we used the derived specific energies and actions (E, J$_{r}$, J$_{\phi}$, J$_{z}$) to cluster them into Chemo-Dynamically Tagged Groups (CDTGs). We then analyzed the elemental-abundance dispersions within these clusters by comparing them to the dispersion of clusters that were generated at random. We find that, for the Group I (primarily CEMP-$s$ and CEMP-$r/s$) clustered stars, there exist statistically insignificant intra-cluster dispersions in [Fe/H], $[\text{C}/\text{Fe}]_{c}$ (evolution corrected carbon), and [Mg/Fe] when compared to the intra-cluster dispersions of randomly clustered Group I CEMP stars.  In contrast, the Group II (primarily CEMP-no) stars exhibit clear similarities in their intra-cluster abundances, with very low, statistically significant, dispersions in $[\text{C}/\text{Fe}]_{c}$. and marginally significant results in [Mg/Fe]. These results strongly indicate that Group I CEMP stars received their carbon enhancements from local phenomena, such as mass transfer from a evolved binary companion in regions with extended star-formation histories, while the CDTGs of Group II CEMP stars formed in low-metallicity environments that had already been enriched in carbon, likely from massive rapidly rotating ultra and hyper metal-poor stars and/or supernovae associated with high-mass early generation stars. 
\end{abstract}


\section{Introduction} 

The nature of the first generation of stars in the Universe, the so-called Population III (Pop III) stars, is of primary interest in Stellar and Galactic Archaeology.  These stars are thought to have been massive and short-lived \citep{Bromm_1999, Omukai_2005}.  Detailed models of rapidly rotating, ultra metal-poor (UMP; [Fe/H] $< -4.0$) and hyper metal-poor (HMP; [Fe/H] $< -5.0$) Pop III stars (e.g., \citealt{Meynet_2006, Meynet_2010, Maeder_2015, Hirano_2015}), as well as the so-called faint (mixing and fallback) supernovae models (e.g., \citealt{Umeda_2005, Nomoto_2013, Tominaga_2014}), which can apply to somewhat higher metallicities, such as extremely metal-poor (EMP; [Fe/H] $< -3.0$) stars, produce prodigious amounts of CNO, capable of enriching the natal gas from which later generations of presently observed, long-lived, low-mass stars are formed.  The initial mass function (IMF) of Pop III stars and the nucleosynthesis pathways involved in their production of the elements incorporated into lower-mass next-generation (Pop II) stars can thus be constrained from studies of carbon-enhanced metal-poor (CEMP) stars, providing essential information on early Galactic chemical evolution and on the nature of the very first stars. 

Carbon can be also produced in significant amounts by low-mass ($M < 1-3 M_{\odot}$) stars in their asymptotic giant branch (AGB) stage of evolution (e.g., \citealt{Suda_2004, Herwig_2005, Lucatello_2005b, Bisterzo_2011, Starkenburg_2014, Hansen_2015a}).  If such a star is a member of a binary (or multiple) system, the transfer of mass to a companion star via Roche lobe overflow (or more likely winds) allows the nucleosynthetic products of the donor (the erstwhile primary star) to be preserved on the surface of the receiving star \citep{Stancliffe_2007,Abate_2015c}. If the receiving star has a mass $M \lesssim 0.7 M_{\odot}$, its lifetime exceeds the Hubble time, and can be observed today.  

The first spectroscopic surveys to assemble significantly large samples of very metal-poor (VMP; [Fe/H] $< -2.0$) Pop II stars in the halo of the Milky Way (MW) were the HK \citep{Beers_1985, Beers_1992} and Hamburg/ESO \citep{Christlieb_2003} objective-prism surveys. \citet{Beers_2005} provided the basis for the modern nomenclature for describing  stars of various (low) metallicities, and suggested initial classifications, based on their carbon and neutron-capture element abundances. Of particular interest to the present study are the various classes of CEMP stars, distinguished by their high level of carbonicity, [C/Fe] (originally set at [C/Fe] $> +1.0$, now usually taken to be [C/Fe] $ >+0.7$).  

Subsequent surveys have greatly enlarged the numbers of recognized CEMP stars to many thousands, including the Sloan Digital Sky Survey \citep{Sloan_2000} and its extensions (SEGUE, \citealt{Yanny_2009}; SEGUE-2, \citealt{Rockosi_2022}), AEGIS \citep{Keller_2007}, LAMOST \citep{Lamost_2012}, and Pristine \citep{PRISTINE_2017}.  

CEMP stars have been demonstrated to increase in frequency with decreasing metallicity. They comprise approximately $30\%$ of all stars with $\text{[Fe/H]} < -2.0$, $60\%$ for $\text{[Fe/H]}<-3.0$, $80\%$ for $\text{[Fe/H]} <-3.5$, and approaching $100\%$ (see \citet{Caffau_2011} for a possible exception) for $\text{[Fe/H]} <-4.0$ (see Figure 6 of \citealt{Yoon_2018}). This fundamental result indicates that the most metal-poor CEMP stars may preserve the 
chemical-abundance signature of the very first generations of stars, making them extremely valuable for the purposes of Stellar and Galactic Archaeology \citep{Frebel_2015a}.

CEMP stars can be divided into multiple sub-classes. Work subsequent to the original \citet{Beers_2005} classification has introduced a number of refinements; we refer the interested reader to the review by \citet{Frebel_2018}. In the present work, we adopt the classifications listed in  Table~\ref{tab:classes}. 
The primary sub-classes of interest in this paper are the CEMP-$s$ stars, which exhibit over-abundances of neutron-capture elements ([Ba/Fe] $> +1.0$, [Ba/Eu] $> +0.5$) associated with production by the $s$-process, and the CEMP-no stars, which exhibit no enhancements of neutron-capture elements ([Ba/Fe] $ < 0.0$)

\begin{deluxetable}{lr}
\tablecaption{\label{tab:classes} CEMP Sub-class Definitions}
\tabletypesize{\scriptsize}
\tablehead{
\colhead{Sub-classes}  & 
\colhead{Definition}}
\startdata
CEMP & [C/Fe] $> +0.7$\\
CEMP-$r$ &  [C/Fe] $> +0.7$, [Eu/Fe] $> +0.7$, [Ba/Eu] $< 0.0$\\
CEMP-$s$ &  [C/Fe] $> +0.7$, [Ba/Fe] $> +1.0$, [Ba/Eu] $> +0.5$\\
CEMP-$i$ ($r/s$) & [C/Fe] $> +0.7$, $0.0 <$ [Ba/Eu] $< +0.5$ or \\
 & [C/Fe] $> +0.7$, $0.0 \leq$ [La/Eu] $\leq +0.6$\\
CEMP-no &  [C/Fe] $> +0.7$, [Ba/Fe] $< 0.0$
\enddata
\end{deluxetable}

It has also been demonstrated that the frequencies of CEMP-$s$ and CEMP-no stars are not the same in different regions (and populations) of the MW.  Specifically, CEMP-$s$ stars are primarily associated with the Metal-Weak Thick Disk (MWTD) and inner-halo regions, while the bulk of the CEMP-no stars are found in the outer-halo region \citep{Carollo_2012, Carollo_2014, Lee_2017, Yoon_2018, Lee_2019}.  These, and other authors, have pointed to this evidence as supporting differences in the formation and accretion histories of the various Galactic stellar populations.

\begin{figure*}[t]
    \includegraphics[width=\textwidth]{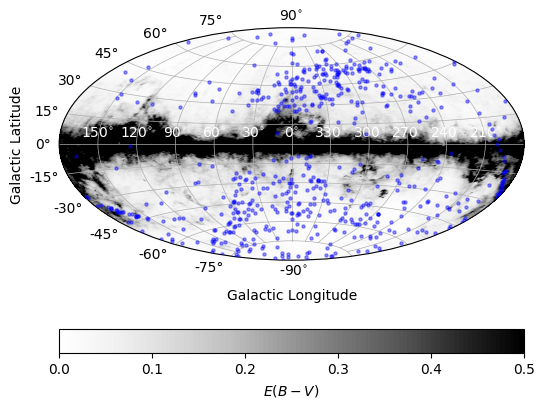}
    \caption{Spatial distribution  of the Initial Sample of CEMP stars. The Galactic reddening map is taken from \citet{Schlegel_1998} and recalibrated by \citet{Schlafly_2011}, shown as the background on a gray scale with darker regions corresponding to greater reddening.}
    \label{fig:gal_map}
\end{figure*}

\citet{Rossi_2005} first pointed out that CEMP stars might not share a single nucleosynthetic origin, based on the apparent bimodal distribution of [C/Fe] and absolute carbon abundance, $A$(C)\footnote{$A$(C) = $\log \epsilon {\rm(C)}$ = $\log (\text{N}_\text{C}/\text{N}_\text{H})+12$, where N$_\text{C}$ and N$_\text{H}$ represent the number densities of carbon and hydrogen, respectively. $A$(C) is interchangeably used as [C/H] + $A$(C)$_\odot$, where $A$(C)$_\odot$ is the absolute Solar carbon abundance.}, at low metallicity (see their Figure 12). Later, \citet{Spite_2013} and \citet{Bonifacio_2015}
presented evidence that there existed high and low ``bands'' in plots of $A$(C) vs. [Fe/H] for main-sequence turnoff stars and mildly evolved subgiants, primarily associated with CEMP-$s$ and CEMP-no stars, respectively. 

\input{Tables/group_stub}

The full richness of the behavior of CEMP stars in the $A$(C)-[Fe/H] space was revealed in Figure 1 of \citet{Yoon_2016} – the so-called Yoon-Beers Diagram, which separated CEMP stars into three morphological groups. Yoon et al. identified the great majority of CEMP-$s$ stars as CEMP Group I stars, based on their distinctively higher $A$(C) compared to the CEMP-no stars, while most CEMP-no stars were classified as either CEMP Group II or Group III stars. The Group II stars exhibit a strong dependency of $A$(C) on [Fe/H], while the Group III stars show no such dependency. Based on this, and the clear differences between Group II and Group III stars in the $A$(Na)-$A$(C) and $A$(Mg)-$A$(C) spaces (Figure 4 of Yoon et al.), these authors argued for the existence of multiple progenitors and/or environments in which the CEMP-no stars formed.

Examination of the effects of dust cooling for varied compositions (e.g., carbon- vs. silicate-based dust grains), could account for the formation of both the Group II and III CEMP-no stars (see, e.g., \citealt{Chiaki_2017}). Simulations of Pop III star enrichment by \citet{Sarmento_2016} exhibited patterns in the $A$(C)-[Fe/H] space that can be associated with these same Groups (see their Figure 13). Simulations of Pop II star formation by \citet{Sharma_2018} show different formation pathways for Group I and Group II CEMP stars. It should be noted that \citet{Yoon_2016} also pointed out that some CEMP stars exhibited anomalous abundance patterns within these different groups, such as stars with $A$(C)$_{c} < 7.1$ while [Ba/Fe] $> 0.0$ or $A$(C)$_{c} > 7.1$ while [Ba/Fe] $< +1.0$. Since the causes of these abundance patterns are not yet understood, their relative numbers are low compared to CEMP stars as a whole, and not all of the CEMP stars have the abundance measurements required to determine their anomalous status, we make no distinctions of these stars in this work.

In light of the above, it is likely that at least some classes of CEMP stars (notably, CEMP-no stars, but also some CEMP-$s$ stars) were not formed in-situ in the Milky Way, but, rather, were born in low- and intermediate-mass satellite dwarf galaxies that were later accreted into it. According to early simulations (e.g., \citealt{de_Zeeuw_2000}), and many since, a large fraction (at least $50\%$) of Galactic accretion events can be recovered by applying clustering algorithms to the phase space of energies and other dynamical parameters of individual field stars. This opens the possibility of using a clustering approach to match CEMP stars formed in similar environments with each other by identifying them as members of individual Chemo-Dynamically Tagged Groups (CDTGs), which we pursue in this paper.

\begin{figure*}[t]
    \includegraphics[width=0.95\textwidth,height=0.95\textheight,keepaspectratio]{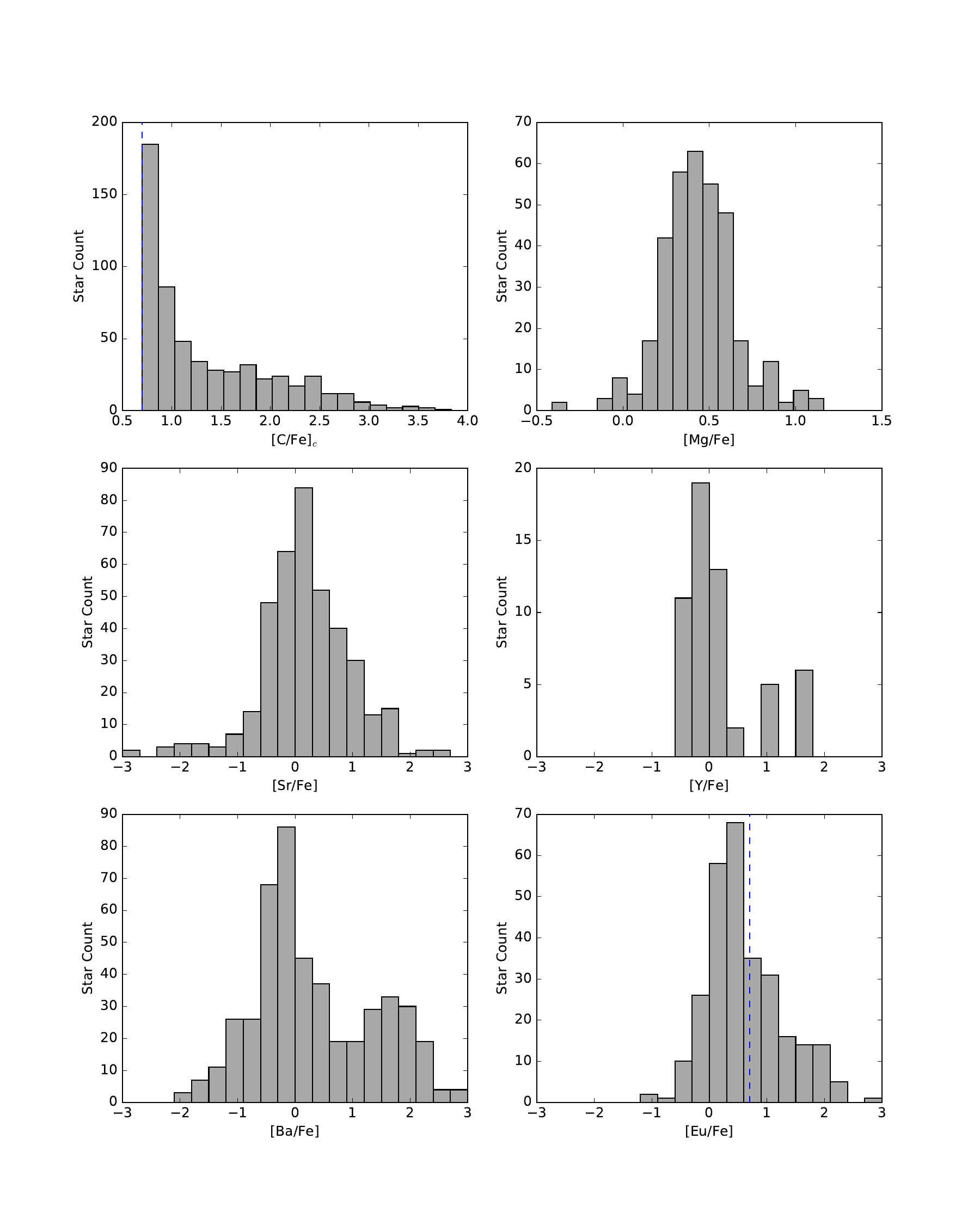}
    \caption{Histograms of the abundances for the Initial Sample of CEMP stars. From left to right, starting at the top, the histograms are as follows: [C/Fe]$_{c}$, [Mg/Fe], [Sr/Fe], [Y/Fe], [Ba/Fe], and [Eu/Fe]. The blue dashed blue line on the the [Eu/Fe] histogram indicates the $r$-II star boundary ([Eu/Fe] $> +0.7$).}
    \label{fig:abundances_histogram}
\end{figure*}

A pilot effort illustrating the application of this approach to chemically peculiar stars in the halo of the MW was conducted by \citet{Roederer_2018}, specific to $r$-process-enhanced stars, based on a relatively small sample size of $35$ such stars. The sample was expanded considerably (426 $r$-process-enhanced stars) by \citet{Gudin_2021}, who demonstrated clear evidence that $r$-process-enhanced stars shared common chemical-evolution histories, presumably in their parent dwarf galaxies. \citet{Shank_2022b} analyzes a sample of  $\sim 1700$ $r$-process-enhanced stars, reaching similar conclusions.  

Another recent work, \citep{Yuan_2020}, explored the clustering of a large data set of VMP MW halo stars. Through application of a trained neural network (STARGO), these authors successfully mapped dynamically tagged groups (DTGs) of VMP stars onto known more-massive sub-structures in the MW, and identified new DTGs for future spectroscopic follow-up. Other recent examples of this approach include an analysis of HK/HES stars \citep{Limberg_2021a}, the Best and Brightest survey \citep{Shank_2022a}, and stars from the RAVE survey \citep{Shank_2022c}.

In this work, we derive dynamical parameters for a sample of $572$ (of an initial sample of $644$) CEMP stars with available moderate- to high-resolution spectroscopy ($R \geq 4350$). 

This paper is outlined as follows.  Section \ref{sec:data} describes the assembly of this sample, along with the adopted distance estimates, radial velocities, proper motions, and the subset of chemical abundances we consider. We also briefly describe the nature of this sample in this section. Section \ref{sec:HDBSCAN} provides a brief overview of the \texttt{HDBSCAN} clustering method we employ, and the results of its application to the CEMP sample. Section \ref{sec:Associations} examines the CDTGs based on the MW substructures and globular clusters that are associated with our CDTGs. In Section \ref{sec:stats} we perform a statistical analysis of our results, and demonstrate the likely shared chemical-enrichment history of the members of the individual CDTGs. We also consider the nature of CDTGs based on clustering of the Group I and Group II CEMP stars in the Yoon-Beers Diagram. %
Section \ref{sec:Discussion} considers the astrophysical implications of our results, and provides perspectives for future studies. Section \ref{sec:Conclusions} summarizes our final results.

\section{Data}\label{sec:data}

A compilation of CEMP stars with abundance analyses based on high-resolution spectroscopy was published by \citet{Yoon_2016}, and it serves as the basis for this work. We also include CEMP surveys with similar analyses -- \citet{Sakari_2018}, \citet{Hansen_2018}, \citet{Ezzeddine_2020}, \citet{Holmbeck_2020}, \citet{Rasmussen_2020}, \citet{Zepeda_2022}, Pristine \citep{Lucchesi_2022}, and GALAH \citep{Buder_2021}. 

Note that we have chosen to exclusively adopt reported abundances based on the local thermodynamical equilibrium (LTE) assumption. Exploration of the effects of the non-LTE assumption, in particular on the abundance estimates for carbon, can and will be undertaken in the future. Popa et al. (2022, A\&A, submitted) have outlined an approach for obtaining non-LTE corrections specifically for the CH molecular feature ($G$-band), upon which the great majority of present high-resolution analyses depend for estimates of carbon.  This is expected to lead to a correction grid covering large ranges of stellar parameters, which will be of general utility.  Note that Popa et al. specifically caution against the naive {\it combination} of non-LTE corrections with 3D effects, as some have attempted in the past; the 3D corrections are still in their relative infancy, and ideally need to be carried out in conjuction with non-LTE assumptions, not separately.

The use of large-scale photometric surveys, e.g., J-PLUS \citep{J-PLUS_2019} and S-PLUS \citep{S-PLUS_2019} will be crucial for identification of many additional relatively bright CEMP stars for future high-resolution spectroscopic follow-up. \citet{Whitten_2019} and \citet{Whitten_2021} have already explored methodology for extracting estimates of [Fe/H] and [C/Fe] from the narrow- and broad-band photometry that they provide. Yang et al. (in preparation) is in the process of extending the elemental-abundance information in these surveys to encompass estimates not only for [Fe/H] and [C/Fe], but for [N/Fe], [Mg/Fe], and [Ca/Fe] as well.  


\begin{figure}[t]
    \includegraphics[width=\linewidth]{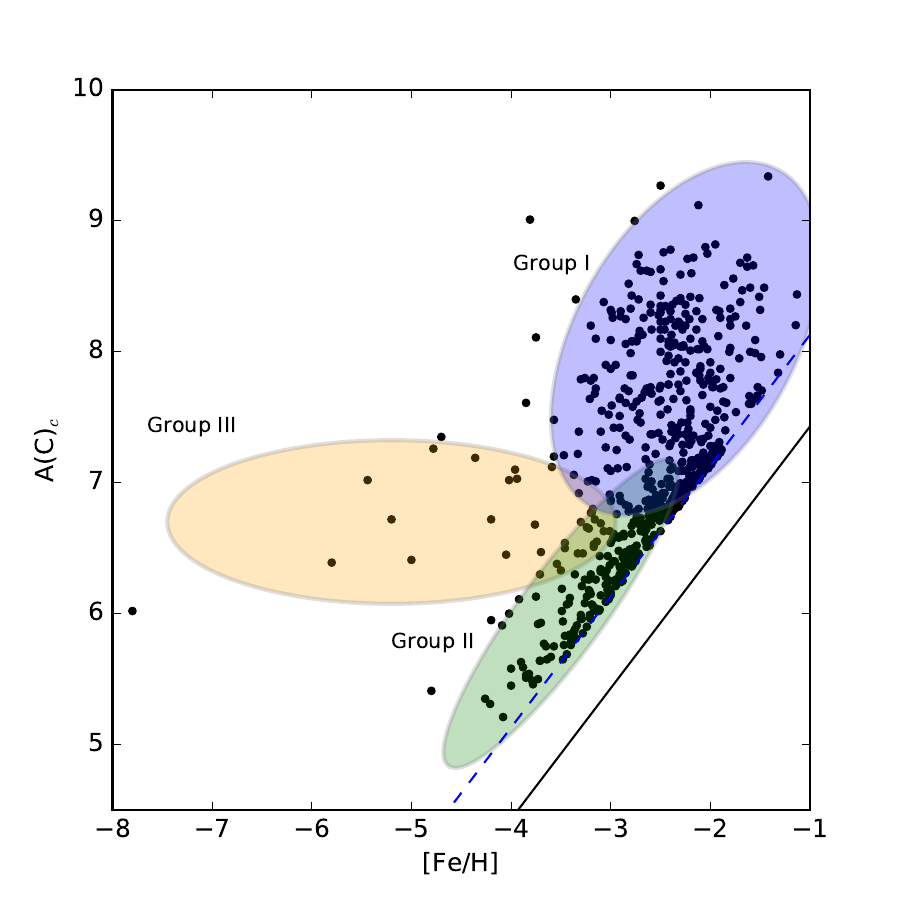}
    \caption{The Yoon-Beers Diagram of the Initial Sample of CEMP stars, showing the evolutionary corrected absolute carbon abundance, $A$(C)$_{c}$, as a function of [Fe/H]. The dotted blue line indicates [C/Fe]$_c = +0.7$, which is the CEMP cutoff. For reference, the solid black line corresponds to [C/Fe]$_c$ = 0.}
    \label{fig:YB}
\end{figure}

\vspace{0.5cm}
\subsection{Construction of the Initial Sample}

The data used in this work was assembled from the literature, including the sources cited in \citet{Yoon_2016}, JINABase \citep{JINABASE_2018}, the SAGA database \citep{Saga}, and a number of other recent sources. We included stars with estimated stellar parameters, [Fe/H], and [C/Fe], at a minimum. When available, we also tabulated the [Mg/Fe] ratio and the neutron-capture elemental-abundance ratios [Sr/Fe], [Y/Fe], [Ba/Fe], and [Eu/Fe].  Duplicates were then removed, retaining the observations of a given star with the higher spectroscopic resolution and additional elemental abundances (in particular $n$-capture species) available. It is unavoidable that the methods used by the different sources are quite inhomogeneous; for this reason we choose not to combine the reported stellar parameter or abundance information when multiple sources exist for a given star.

\begin{figure}[t]
    \includegraphics[width=\linewidth]{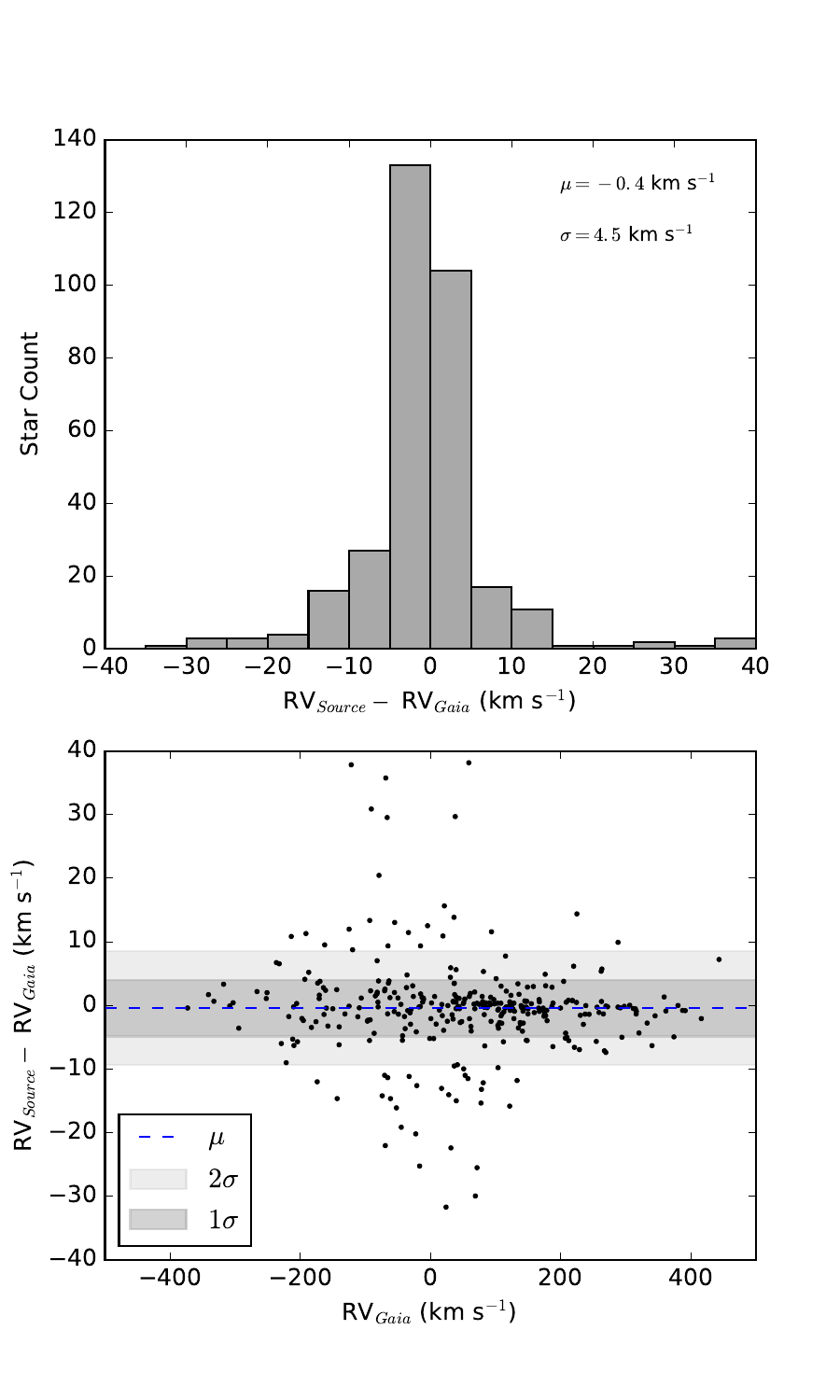}
    \caption{Top Panel: Histogram of the differences between the Source and $Gaia$ values of radial velocity. Bottom Panel: The residuals between the Source and $Gaia$ values of radial velocities, as a function of the $Gaia$ radial velocity. The dotted blue line shows the mean of the residuals ($-0.4$ km s$^{-1}$). The dark and light shaded regions correspond to 1$\sigma$  ($4.5$ km s$^{-1}$) and 2$\sigma$  ($9.0$ km s$^{-1}$) error in the residuals, respectively.}
    \label{fig:rv_diff}
\end{figure}

\begin{figure}[t]
    \includegraphics[width=\linewidth]{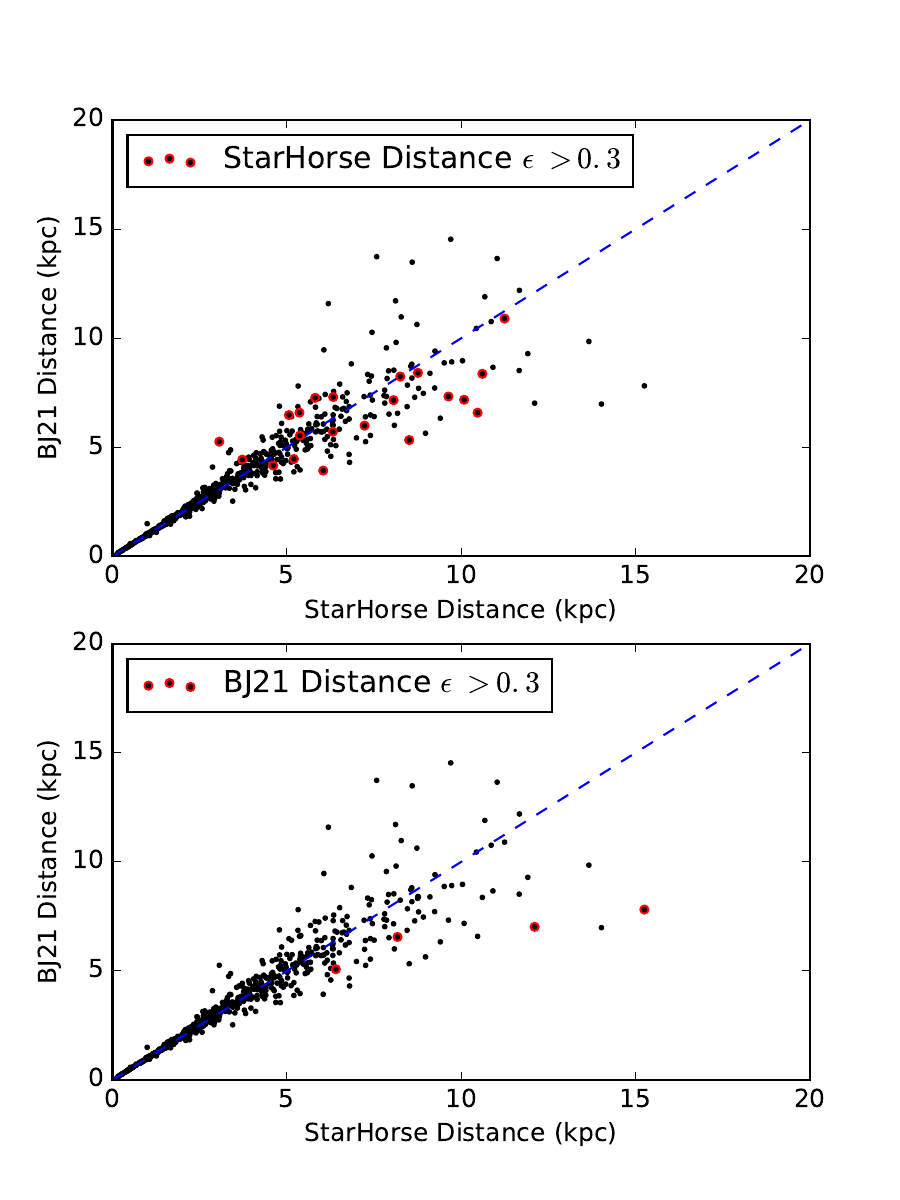}
    \caption{A comparison of the distance estimates from StarHorse \citep{Anders_2021} and Bailer-Jones \citep{Bailer_Jones_2021} for stars in the Initial Sample. Top Panel: Bailer-Jones distance as a function of StarHorse distance with stars with StarHorse relative distance errors $\epsilon >  30\%$ highlighted in red. Bottom Panel: Bailer-Jones distance as a function of StarHorse distance with stars with Bailer-Jones relative distance errors $\epsilon > 30\%$ highlighted in red. The blue-dashed line in both panels indicates a one-to-one comparison.}
    \label{fig:distance}
\end{figure}

\begin{figure*}[t!]
    \includegraphics[width=\textwidth,height=0.4\textheight]{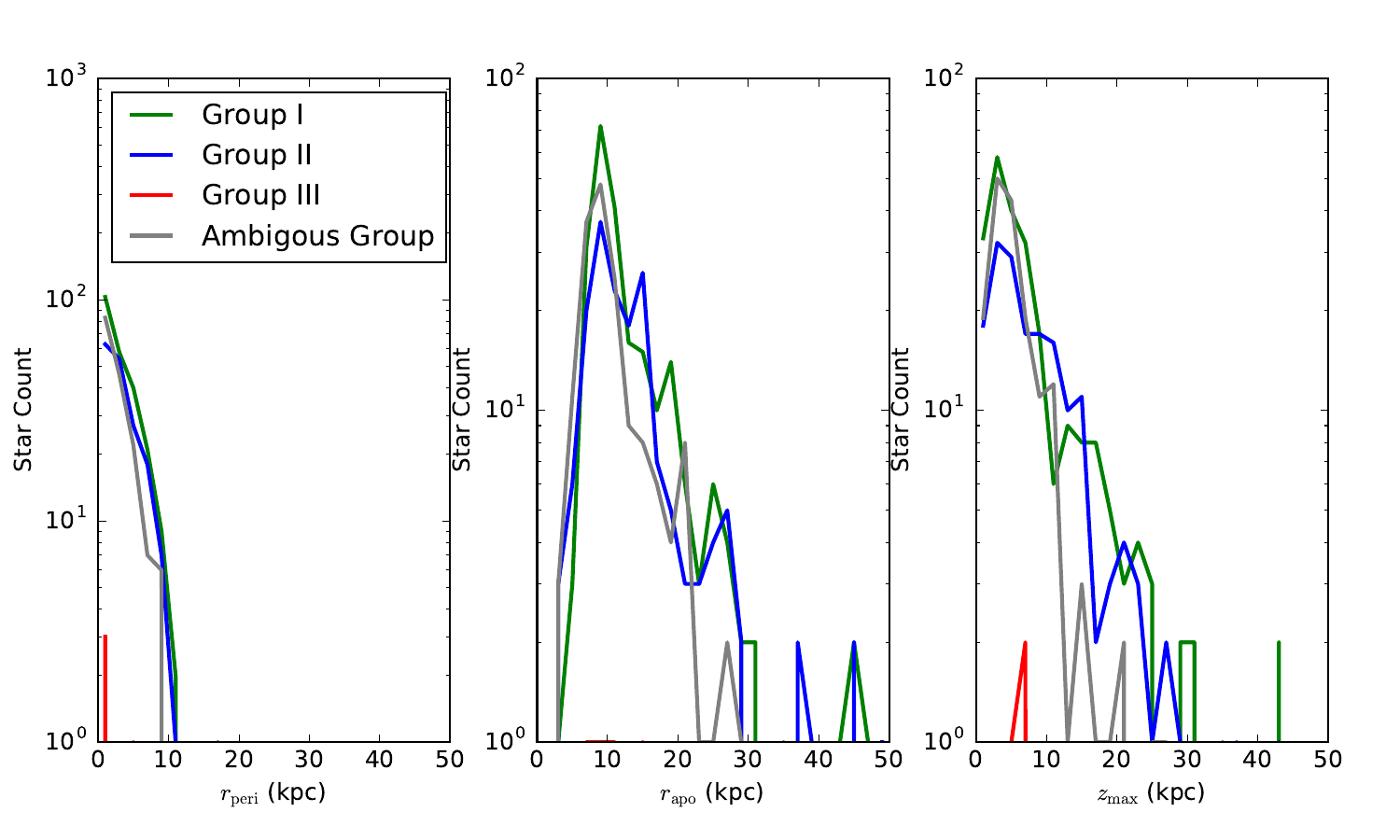}
    \caption{Distributions of $r_{\text{peri}}$, $r_{\text{apo}}$, and Z$_{\text{max}}$, from left to right. The stars are separated into their Yoon-Beers CEMP morphological groups, as indicated in the left panel. Stars that were unbound from the Galaxy according to the \texttt{AGAMA} results are not included.}
    \label{fig:ord_dist}
\end{figure*}

\begin{figure*}[t]
    \includegraphics[width=\textwidth,height=0.4\textheight]{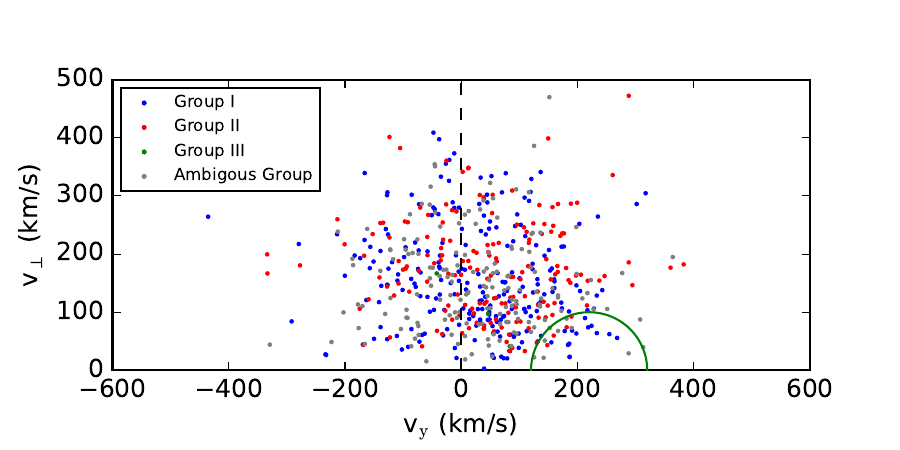}
    \caption{The Toomre Diagram of the Final Sample. The axes are v$_{\perp} = \sqrt{\text{v}_{x}^{2} + \text{v}_{z}^{2}}$ vs. v$_{y}$. The green circle represents the location of $100$ km s$^{-1}$ from the Local Standard of Rest ($232$ km s$^{-1}$), while the vertical black dashed line represents the division between prograde (v$_{y} > 0$ km s$^{-1}$) and retrograde (v$_{y} < 0$ km s$^{-1}$) stellar orbits. The Group I stars are indicated as blue points, Group II stars are indicated as red points, Group III stars are indicated as green points, and the stars with ambiguous Group assignments are plotted in gray.}
    \label{fig:toomre}
\end{figure*}

We then removed stars that did not satisfy [C/Fe]$_{c} > +0.7$, obtained by applying the carbon-abundance correction scheme for evolution on the giant branch \citep{Placco_2014}. Following these cuts, our Initial Sample consists of 644 stars. This sample includes stars with  {$3540  \leq T_{\text{eff}} \rm(K) \leq 7100$, surface gravity $-0.30 \leq  \log g \leq 5.07$, $-7.80 \leq {\rm [Fe/H]} \leq -1.13$, and $+0.71 \leq {\rm [C/Fe]}_{c} \leq +4.39$}. The spatial distribution of the stars in our Initial Sample of CEMP stars are shown in Figure \ref{fig:gal_map}. Table \ref{tab:initial_data_descript} in the Appendix lists the various data for these stars. In the print edition, only the table description is provided; the full table is available only in electronic form.

Figure \ref{fig:abundances_histogram} provides histograms of the full set of elemental-abundance ratios considered in the present work.

The Yoon-Beers Diagram of $A(\rm{C})_{c}$ vs. [Fe/H] for these stars is shown in Figure~\ref{fig:YB},
corrected from the observed value to account for the depletion of carbon on the giant branch following \citet{Placco_2014}. The definitions of CEMP morphological groups were originally presented in \citet{Yoon_2016}, based on the level of [Ba/Fe]. These same authors demonstrated that similar assignments can be carried out using a separation based on $A(\rm{C})_{c}$ (and [Fe/H]) alone. As not all of our CEMP stars have available measured [Ba/Fe], we base our morphological group assignments purely on corrected carbon abundances, $A(\rm{C})_{c}$, [C/Fe]$_c$, and [Fe/H]. To account for the errors in these measurements, and the lack of clear delineations between the groups, we use definitions with some overlap. We assign CEMP groups as follows: Stars with [Fe/H] $> -4$ and $A(\rm{C})_{c}$ $> 6.75$ are Group I stars, stars with $A(\rm{C})_{c}$ $< 7.25$ and [C/Fe]$_{c}$ $< +2.25$ are Group II stars, and all other stars are assigned to Group III.  With these definitions, we have 443 Group I stars (271 Group I only stars), 363 Group II stars (191 Group II only stars), and 10 Group III stars. For convenience, a list of the stars with their adopted CEMP Group status and a summary of their available elemental-abundance ratios is provided in Table~\ref{tab:group_stub}. Note that this table includes all stars in the Initial Sample of CEMP stars. Those that are removed in the assembly of the Final Sample, as described below, are listed with an asterisk following their names.

\subsection{Construction of the Final Sample}

\subsubsection{Radial Velocities, Distances, and Proper Motions}

To obtain the dynamical information for our Initial Sample we cross-matched our stars with $Gaia$ EDR3 \citep{Gaia_2016, Gaia_2021}, supplemented by recent radial velocity information from $Gaia$ DR3 \citep{Gaia3}, in order to obtain radial velocities, parallaxes, and proper motions. When an accurate radial velocity was available from the spectroscopic surveys (which applies to all but a handful of cases), we use it if there was no $Gaia$ RV.  For stars with available Source radial velocities, Figure \ref{fig:rv_diff} compares these values with those having $Gaia$ radial velocities. If the spectroscopic Source radial velocities differed from those reported by $Gaia$ by more than 15 km~s$^{-1}$ (45 stars), the star was removed from further analysis, as many are suspected to be binaries.  

Distance estimates for our stars were obtained using StarHorse \citep{Anders_2021}. When the relative error from StarHorse exceeded $30\%$, we used the Bailer-Jones \citep{Bailer_Jones_2021} estimates, unless its relative error was also greater than $30\%$, in which case we removed the star from the sample. Figure \ref{fig:distance} shows the distance estimates from both methods for stars in the Initial Sample. Note that three stars with unusually high, and likely suspect, distances were removed from the Final Sample.

\subsubsection{Dynamical Parameters}

We employ the Action-based GAlaxy Modelling Architecture\footnote{\url{http://github.com/GalacticDynamics-Oxford/Agama}} (\texttt{AGAMA}) package \citep{Vasiliev_2018}, which uses the radial velocities, distances, and proper motions of stars to derive orbital parameters for stars in the Initial sample. We use the same Solar position, Solar peculiar motions\footnote{We adopt a Solar position of ($-8.249$, $0.0$, $0.0$) kpc \citep{GravityCollaboration2020} and Solar peculiar motion in (U,W) as ($11.1$,$7.25$) km s$^{-1}$ \citep{Schonrich_2010}, with Galoctocentric Solar azimuthal velocity  \textit{V} $= 250.70$ km s$^{-1}$ determined from \citet{Reid_2020}.}, and gravitational potential \citep{McMillan_2017} as used in \citet{Shank_2022a}. To obtain errors for these estimates, we assume the errors in the input quantities are normally distributed. Then we obtain orbital parameters by finding the mean and standard deviation of the values from 1,000 runs of \texttt{AGAMA} sampling from the distributions for our inputs. 

Figure \ref{fig:ord_dist} presents histograms of the derived estimates of $r_{\rm peri}$, $r_{\rm apo}$, and Z$_{\rm max}$; stars that were unbound from the Galaxy (E $> 0$ km$^{2}$ s$^{-2}$) are not included. All of the stars have $r_{\rm peri}$ distances inside of 20 kpc. 
As can be appreciated from inspection, the distributions of these parameters are quite similar across the Group I and Group II sub-samples. There are too few Group III stars to make meaningful interpretations at present.


The stars with \texttt{AGAMA} results consistent with being bound to the Galaxy comprise our Final Sample, which includes a total of 572 stars. Table \ref{tab:final_data_descript} in the Appendix lists the various data for these stars. In the print edition, only the table description is provided; the full table is available only in electronic form.

Figure \ref{fig:toomre} is a Toomre Diagram of the stars in the Final Sample. For this sample, $64\%$ of the stars are on prograde orbits, while $36\%$ are on retrograde orbits. When we consider the stars that are assigned (unique) morphological groups, Group I stars are $62\%/38\%$ prograde/retrograde. Group II stars are $68\%/32\%$ prograde/retrograde. The Group III stars are $75\%/25\%$ prograde/retrograde. Note that not all stars in the Final Sample have morphological groups assigned, so these numbers are based on the subset that do. 



\input{Tables/cluster_summary_table}

\section{Clustering Procedure}\label{sec:HDBSCAN}


\input{Tables/cluster_stellar_results_stub_table}

To perform our clustering exercise, we employ  \texttt{HDBSCAN}\footnote{For a detailed description of the \texttt{HDBSCAN} algorithm visit: \url{https://hdbscan.readthedocs.io/en/latest/how_hdbscan_works.html}} \citep{Campello_2013}. This method groups data by density in the phase space considered (orbital energies and cylindrical actions), then places the clusters into a hierarchy based on how they change when the requisite density belonging to a cluster changes. We refer the interested reader to  \citet{Campello_2013}, and \citet{Shank_2022a} for a full description of \texttt{HDBSCAN}. In our use of this algorithm, we set the following parameters: \verb ~min_cluster_size ~$= 3$, \verb ~min_samples ~$= 3$, \verb ~cluster_selection_method ~$=$ \verb 'leaf' , \verb ~prediction_data ~$=$ \verb ~True~, Monte Carlo samples set at $1000$, and minimum confidence set to $20\%$. The \verb ~min_cluster_size~ determines how small the clusters can be. The \verb ~min_samples~ parameter can be adjusted to account for different noise levels in the data. By choosing \verb ~cluster_selection_method ~$=$ \verb 'leaf'~ we allow \texttt{HDBSCAN} to make smaller clusters with tighter orbital parameter values. With the \verb ~prediction_data ~$=$ \verb ~True~ the method has memory of the nominal clusters for each run in the Monte Carlo procedure. Full descriptions of these input choices of parameters can be found in \citet{Shank_2022a}.

Table \ref{tab:cluster_summary} lists the $40$ identified CDTGs, along with the number of members, assigned confidence values, and associations to previously identified DTGs and CDTGs, as described below.  Note that, even though we set the minimum confidence for cluster identification to $20.0\%$, the smallest confidence in this table is $35.5\%$; most are much higher. The average confidence level for the $40$ CDTGs is $77.1\%$.

Table \ref{tab:cluster_results_stub} provides a listing of the individual members of the identified CDTGs from the application of \texttt{HDBSCAN}, along with the available elemental-abundance ratios considered in this paper. The biweight location and scale, analogous to the mean and standard deviation, for the abundances within each CDTG are shown in bold at the end of each listing. In addition, any stars that are associated with DTGs, CDTGs, globular clusters, or dwarf galaxies identified in previous works are listed at the beginning of each CDTG in the table.

\input{Tables/cluster_orbital_table}

Table \ref{tab:cluster_orbit} lists the mean values, along with the associated dispersions, of the dynamical parameters for the CDTGs. The number of stars in each CDTG is indicated by the $N$ Stars column.
%

\section{Structure Associations}\label{sec:Associations}

\subsection{Milky Way Substructures}
\begin{figure*}[t!]
\centering
    \includegraphics[width=0.89\textwidth,height=0.89\textheight,keepaspectratio]{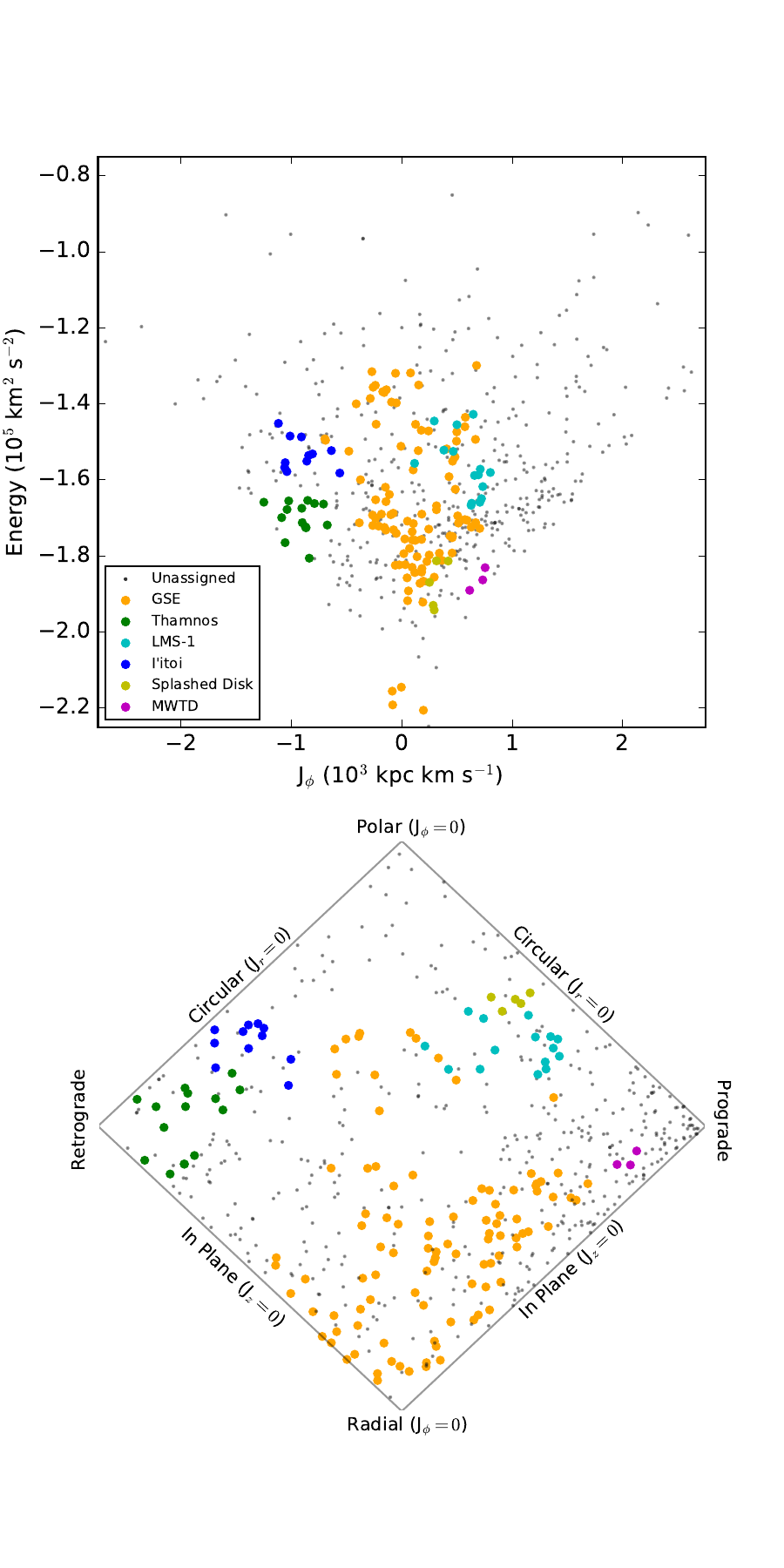}
    \caption{Locations of the CEMP CDTG stars in two orbital-parameter phase spaces. Top Panel: Lindblad Diagram of the identified MW substructures. The different structures are associated with the colors outlined in the legend. The gray points are the stars from the Final Sample that are not assigned to any CDTGs. Bottom Panel: The projected-action plot of the same substructures. This space is represented by J$_{\phi}$/J$_{\text{Tot}}$ for the horizontal axis and (J$_{\text{z}}$ - J$_{\text{r}}$)/J$_{\text{Tot}}$ for the vertical axis with J$_{\text{Tot}}$ = J$_{\text{r}}$ + $|$J$_{\phi}|$ + J$_{\text{z}}$. For more details on the projected-action space, see Figure~3.25 in \citet{Binney2008}.}
    \label{fig:cluster}
\end{figure*}

We check each CDTG for association with known MW substructures. These associations are based on the stellar orbital parameters as well as the chemical abundances of individual stars in the CDTGs. To find these associations we employ the substructure criteria from \citet{Naidu_2020}. Each substructure has specific dynamical and chemical criteria, and CDTGs are associated to these substructures when the criteria are met. To see exactly how these criteria are applied, we refer the interested reader to \citet{Shank_2022b}. We found that $25$ of the $40$ CTDGs had associated substructures: Gaia-Sausage-Enceladus (GSE), LMS-1 (Wukong), the Metal-Weak Thick Disk (MWTD), Thamnos, I'itoi, and the Splashed Disk listed in Table~\ref{tab:substructures}. This table provides the numbers of stars in the CDTGs associated with each substructure, the means and dispersions of their chemical abundances, and the means and dispersions of their dynamical parameters. The Lindblad Diagram and projected-action plot for these substructures is shown in Figure~\ref{fig:cluster}. The CDTGs found to be associated with known MW substructures and globular clusters are listed in Table \ref{tab:interesting_substructure}.

\clearpage

\subsubsection{Gaia-Sausage-Enceladus}\label{subsubsec:GSE}
One of the earliest known mergers, Gaia-Sausage-Enceladus (GSE), is thought to have been accreted by the MW about 10 Gyr ago \citep{Helmi_2020}. It is also the largest known merger, and the most populated substructure in this work, with 99 CEMP stars in 18 different CDTGs. 
While the metallicity distribution function for GSE peaks near [Fe/H] $\sim$ $-1.2$, the CEMP stars associated with GSE exhibit a biweight location and scale (robust alternatives to the mean and standard deviation) of $\langle$[Fe/H]$\rangle = -2.52$ ($ \sigma = 0.55$ dex). Our GSE stars also show a biweight location and scale of $\langle$[Mg/Fe]$\rangle = +0.40$ ($ \sigma = 0.22$ dex), which is higher than GSE's value ($+0.21$;
\citealt{Naidu_2022}). Based on these differences, it is likely that the low-metallicity tail of GSE is comprised of CEMP stars with elevated [Mg/Fe] abundances. 

\subsubsection{LMS-1 (Wukong)}\label{subsubsec:LMS1}
We found 2 CDTGs in the substructure LMS-1, with a total of 15 stars. LMS-1 was found to be the most metal-poor substructure by \citet{Malhan_2022}, so finding this large number of CEMP stars might be expected, given the increasing frequency of CEMP stars with declining metallicity. The CEMP stars associated with LMS-1 belong to all three CEMP Group morphologies. We found 3 UMP ([Fe/H] $< -4$) stars associated with LMS-1. No other substructure has more than 2 stars at or below [Fe/H] $< -4$, and there was only one such star in a CDTG without an associated substructure. Based on this, UMP stars are likely much more common in LMS-1 than in other recognized MW substructures, despite the mean of $-1.58$ reported by \citet{Naidu_2022} for the substructure as a whole. For our sample, we found $\langle$[Fe/H]$\rangle = -3.11$ with a large dispersion of $\sigma = 0.78$ dex.

\input{Tables/substructure_table}

\subsubsection{Thamnos}\label{subsubsec:Thamnos}
Thamnos is believed to have formed from the merging of a relatively small satellite (stellar mass $< 5 \times 10^{6}\,M_{\odot}$) with the MW \citep{Helmi_2020}. It is particularly interesting how low in energy the stars in this structure are, considering how deep in the MW it resides. This has been used to argue that the merger occurred very early in the MW's history \citep{Bonaca_2020,Kruijssen_2020}. We found 2 CDTGs associated with Thamnos, with 15 total CEMP stars, including both Group I and Group II CEMP morphologies. The stars in Thamnos have the second highest mean metallicity ($\langle$[Fe/H]$\rangle = -2.35$) among those with substructure associations, with a dispersion $\sigma = 0.30$ dex. 

\subsubsection{The Metal-Weak Thick Disk}\label{subsubsec:MWTD}
The MWTD has been argued to have formed from either a merger scenario, possibly related to GSE, or the result of old stars born within the Solar radius migrating out to the Solar position due to tidal instabilities within the MW \citep{Carollo_2019}. However, several recent papers, including \citet{Carollo_2019}, \citet{An2020}, \citet{Dietz_2021}, and \citet{Mardini_2022}, have presented evidence that the MWTD is an independent structure from the canonical thick disk, and thus may have arisen independently.

We found 1 CDTG of CEMP stars associated with this substructure, with a combined total of 3 stars. As expected, these stars occupy the low-metallicity tail of the MWTD Metallicity Distribution Function (MDF), with a biweight location and scale of $\langle$[Fe/H]$\rangle = -2.28$ ($\sigma = 0.20$ dex). The CEMP stars associated with the MWTD have a [Mg/Fe] biweight location and scale ($\langle$[Mg/Fe]$\rangle = +0.36$, $\sigma = 0.01$ dex) that is quite similar to the stars in GSE.

\subsubsection{I'itoi}\label{subsubsec:I'itoi}
The substructure I'itoi has several possible formation scenarios suggested in the literature, associating it with other halo substructures, including Sequoia and Thamnos \citep{Naidu_2020}. We found 1 CDTG in I'itoi that contains a total of 11 stars, including both Group I and Group II morphologies. We found I'itoi to be the second most metal-poor substructure among our associations,  $\langle$[Fe/H]$\rangle = -2.71$) with a dispersion of $\sigma = 0.33$
dex.

\subsection{Splashed Disk}

The Splashed Disk (SD) is the second least-populated substructure found to be associated with our CEMP CDTGs. It is believed to have been formed when a component of the primordial disk was heated by the GSE merger, giving it its now characteristic kinematics \citep{Naidu_2020}. The one CDTG associated with this substructure is unique for the fact the [C/Fe$]_{c}$ and [Mg/Fe] ratios have larger dispersions than any other substructures. and the [Fe/H] dispersions is the second largest of all the substructures. We note this as interesting, but don't have an explanation for why this might be besides the small statistics due to the limited number of stars.  

\input{Tables/interesting_substructure_table}

\subsection{Previously Identified Dynamically Tagged Groups and Stellar Associations}\label{subsec:Prev_DTGs_Stellar_assoc}
To further validate the results of these cluster associations with MW substructures, we employ the mean group orbital properties of each CDTG to compare with previously identified CDTGs. We then make associations between stars in our CDTGs with previously DTGs, based on angular separation on the sky. We consider stars as being associated when the angular separation is within $5 \arcsec$. 

The full list of CDTGs associated with DTGs, groups, streams, or populations from other works is shown in Table \ref{tab:interesting_groups_substructure}.  For example, for CDTG-1 we found no substructure associations, but found it to be associated with DG21:CDTG-3 and DS22b:DTG-51 through the group mean orbital parameters, and found star-to-star associations with both of the aforementioned DTGs from \citet{Gudin_2021, Shank_2022b}. While both CDTG-1 and DS22b:DTG-51 did not have associations with a MW substructure, DG21:CDTG-3 was associated with the MWTD. Based on the fact that CDTG-1 has 17 member stars, DS22b:DTG-51 has 11 member stars, and DG21:CDTG-3 only has 3 member stars, the stars in CDTG-1 are not likely connected to the MWTD, as DG21:CDTG-21 suggested.

If we examine the one of the most metal-poor CDTGs identified, CDTG-15, associated with LMS-1, we find that one member, HE~1310-0536, was inferred as being part of the MW outer halo by \citet{Sestito_2019}. This demonstrates the ability to find outer-halo substructures that are currently near their $r_{\text{peri}}$, and thus more easily observable.

Another use for associating our CDTGs with previous studies is the ability to examine substructures with few associated CDTGs. We found only CDTG-3 associated with the I'itoi substructure. However, there are 2 CEMP stars with moderate $r$-process enhancement. \citet{Naidu_2020} has argued that I'itoi, and the larger substructure it is a part of, forms the dominant component of the highly retrograde halo. This suggests that mergers associated with the highly retrograde halo may have preferentially experienced star-formation environments undergoing $r$-process enrichment. Clearly, this association of stars with a characteristic CEMP-$r$ abundance signature is worthy of further study.  

\subsection{Globular Clusters and Dwarf Galaxies}\label{subsec:GCDG}
In previous work using this methodology to find associations between CDTGs, DTGs, and MW substructures, globular clusters have frequently been identified as possibly linked to the dynamical clusters. However, of the 40 CDTGs found in this work, only 4 exhibit associations with globular clusters: CDTG-6, CDTG-20, CDTG-32, and CDTG-40. The first and second of these CDTGs were found to be associated with Thamnos and GSE, respectively (the last two had no associated substructure). It is certainly of interest how few globular clusters can be associated with CEMP CDTGs, and worthy of investigation. CEMP stars have been shown by a number of authors to be rare in globular clusters (e.g., \citealt{Kirby_2015, Arentsen_2021}, and references therein). It has been suggested that the lack of CEMP-$s$ (Group I) stars in globular clusters is due to the low rate of long-lived binary stars in these dense environments. The present work clearly supports this hypothesis.  If CEMP-no (Group II) stars formed early, from gas enriched by the ejecta of high-mass supernovae 
(SNe), as we argue in this paper, their rarity in presently observed globular clusters might be expected. 

\section{Chemical Structure of the Identified CDTGs}\label{sec:stats}

\subsection{Statistical Framework}

Following \citet{Gudin_2021}, we perform a statistical analysis of our CDTGs to determine how probable the observed abundance dispersions for a given set of elements would be if their member stars were selected at random from the full set of CEMP stars in the Final Sample. To perform this analysis, we create $2.5\times 10^6$ random groups of $3 \leq N \leq 12$ stars with dispersions based on the biweight scale \citep{Beers_1990}. We then use this to generate cumulative distribution functions (CDFs) for the abundance distributions for each possible size of the random groups. This serves as a proxy for abundances we should expect in a CDTG of a given size. If, for a given element, the CDTGs preferentially populate the low end of the simulated CDFs, we infer that its members exhibit strong similarities in their abundances. We then use binomial and multinominal probabilities to calculate the statistical likelihoods of the distributions of the CDTG abundance dispersions based on the derived CDFs.

We calculate the statistical significance at three thresholds of CDF values ($\alpha \in \{0.25, 0.33, 0.5\}$) for each of the individual elemental abundances ($X\in \{\text{[Fe/H]}, \text{[C/Fe]}_\text{c}, \dots \}$) of the CDTGs, as well the significance across all $\alpha$ values, or across all X abundances. These probabilities are defined as:

\begin{itemize}
    \item \textit{Individual Elemental-Abundance Dispersion (IEAD) probability}: Individual binomial probability for specific values of $\alpha$ and $X$.
    \item \textit{Full Elemental-Abundance Distribution (FEAD) probability}: Multinomial probability for specific values of $\alpha$, grouped over all abundances $X$.    
    \item \textit{Global Element Abundance Dispersion (GEAD) probability}: Multinomial probability for specific abundances $X$, grouped over all values of $\alpha$. This is the overall statistical significance for the particular abundance.
    \item \textit{Overall Element Abundance Dispersion (OEAD) probability}: Multinomial probability grouped over all values of $\alpha$ and all abundances $X$. This is the overall statistical significance of our clustering results.
\end{itemize}

For a more detailed discussion of the above probabilities, and their use, the interested reader is referred to \citet{Gudin_2021}.

\subsection{Important Caveats}

It should be noted this statistical analysis method requires that the elemental abundances of the parent sample of stars, the population that the clustering was performed on, has dispersions that are large enough that the individual CDTG elemental dispersions are sufficiently low in comparison.  If the dispersions of the parent population are too small compared to the CDTGs for that element, the analysis  will not be able to find distinctions between a random grouping of stars drawn from the parent population and the stars within a given CDTG. To aid in determining which elements this applies, we compare the Inter-Quartile Range (IQR) of the means of random draws from from the parent sample to the IQR of the means of the CDTGs for a given element. We adopt the rule of thumb that the mean IQR for each element of a CDTG is on the order of one-half of the IQR of the randomly drawn stars from the parent population.  Otherwise, there is insufficient ``dynamical range" for the statistical inferences to be made with confidence, at least for individual elements.

\vspace{2cm}

\input{Tables/interesting_substructure_groups_table}

The statistical power of our comparisons increase with the numbers of CDTGs in a given parent population. This is ultimately the reason that we set the minimum number of stars per CDTG to 3, so that when we examine the sample divided into CEMP Group I and Group II morphologies, as described below, we retain a sufficiently large number of CDTGs (in particular for the Group II CEMPs) for meaningful statistical comparisons. 

It should also be kept in mind that the observed CDTG elemental-abundance dispersions depend on a number of different parameters, including not only the total mass of a given parent dwarf galaxy, but on its available gas mass for conversion into stars, the history of star formation in that environment, and the nature of the progenitor population(s) involved in the production of a given element. These are complex and interacting sets of conditions, and certainly are best considered in the context of simulations (such as \citealt{Hirai_2022}). Consequently, the expected result for a given element in a given set of CDTGs is not always clear. However, we have designed our statistical tests to consider a broad set of questions of interest, the most pertinent of which for the current application are the FEAD and OEAD probabilities, which we employ for making our primary inferences. 

\clearpage

\input{Tables/cluster_mean_table}

\input{Tables/binomial_probability_table_full}

\subsection{Results}

\subsubsection{The Full Sample of CEMP CDTGs}

Table \ref{tab:cluster_mean} lists the means and dispersions of the elemental abundances explored in this study for each of the CDTGs identified in this work. The second part of the table lists the global CDTG properties, with the mean and standard error of the mean (using biweight location and scale) of both the CDTG means and dispersions being listed. The second part of the table also includes the IQR (Interquartile Range) of the CDTG means and dispersions. The third part of the table lists the biweight location and scale of the elemental abundances in the Final Sample, along with the IQR of the elemental abundances in the Final Sample. As can be seen by comparing the two IQRs for each the CDTG results and the Final Sample (indicated in bold face), the IQRs for the CDTG results for only 1 of the 7 elements considered is at least twice smaller ([C/Fe]$_c$), while the remaining 5 of the 7 elements have IQRs that narrowly miss the our rule of thumb, the exception being [Y/Fe], which has relatively few CDTGs with available dispersions. Going forward, we choose to only consider the statistical inferences that can be drawn by considering three elements, [Fe/H], [C/Fe]$_c$, and [Mg/Fe], and set the neutron-capture elements aside for our present analysis. We remind the reader that we are comparing how individual CDTGs dispersions compare to randomly generated clusters.

Table \ref{tab:binomial_probability_table_full} lists the numbers of CDTGs with available estimates of the listed abundance ratios for [Fe/H], [C/Fe]$_c$, and [Mg/Fe], the numbers of CDTGs falling below the $0.50$, $0.33$, and $0.25$ levels of the CDFs, and our calculated values for the various probabilities. Both the full and the overall probabilities (captured by the FEAD and the OEAD probability values) are statistically significant, indicating a similarity for all of the considered elements within our CDTGs across the entire sample (the only exception being the FEAD of $N<0.5$). The individual abundance spreads (the GEAD probabilities), vary from marginal statistical significance for [Fe/H] to a lack of significance for [C/Fe]$_c$ and [Mg/Fe]. Keep in mind, as noted above, that the contrast in the mean IQRs of all three elements for our CDTGs with their can IQRs for the full sample may impact interpretation of their probabilities.


There are numerous lines of evidence that point towards Group I CEMP stars receiving their carbon enhancement from local events, for example accretion of material from a binary companion. This would clearly not affect the other stars in a given CDTG, and could vary a great deal even across the subset of CEMP Group I stars in an individual CDTG, given the ranges of mass possible for their erstwhile primary members, the initial separations of their members, and a host of other variables. To investigate this idea further, we proceed by separating the CEMP stars into their morphological groups, after removal of stars not assigned to groups, or which had ambiguous group assignments.  We then carried out dynamical clustering within these subsets, and then performed the statistical analysis on each subset. 

\subsubsection{The CEMP Group I and CEMP Group II Stars}

The CDTGs of CEMP Group I stars and CEMP Group II stars are listed in Table \ref{tab:g1_cluster_results_stub} and Table \ref{tab:g2_cluster_results_stub} of the Appendix, respectively. From inspection of Table \ref{tab:binomial_probability_table_full}, the FEAD and GEAD probabilities are substantially higher for the CDTGs of Group I CEMP stars than found for the Full Sample, and only the OEAD probability ($12.0\%$) approaches marginal significance. However, for the Group II CEMP stars, the FEAD probabilities are either nearly significant (5.6\% in the case of $N < 0.50$), or highly significant.  The GEAD probability for [Fe/H] is not significant, while that for [C/Fe]$_c$ is highly significant, and for [Mg/Fe] is marginally significant.



Inspection of Tables \ref{tab:g1_cluster_mean} and \ref{tab:g2_cluster_mean} in the Appendix underscore these differences. While the dispersions of the mean [Fe/H] for both the Group I and Group II CDTGs are essentially identical (0.30 dex and 0.21 dex, respectively), the contrast in the dispersions of the mean [C/Fe$]_{c}$ between the Group I and Group II CDTGs are quite clear (0.47 dex and 0.10 dex, respectively).  Indeed, the  {\it highest} [C/Fe]$_c$ dispersion of the Group II CDTGs is similar to the {\it mean} Group I CDTG [C/Fe]$_c$ dispersion. 

Table \ref{tab:g1_cluster_mean} shows that the IQRs of the mean [Fe/H] and [C/Fe]$_c$ for the Group I CDTGs meets our factor of two rule of thumb compared to the parent population draws for these elements, while that for [Mg/Fe] (0.31 dex) is essentially identical to that for its parent population (0.26 dex). Table \ref{tab:g2_cluster_mean} shows that the IQRs of the Group II stars for [Fe/H], [C/Fe]$_c$, and [Mg/Fe] all satisfy our rule of thumb.

\section{Discussion and Future Perspectives}\label{sec:Discussion}

We now consider how to account for the observed behaviors of the CDTGs in our study, emphasizing the likely differences in the astrophysical origins of Group I (primarily CEMP-$s$) stars and
Group II (primarily CEMP-no) stars.

Numerous studies support the binary mass-transfer scenario for the production of carbon for Group I stars. The work of \citet{Lucatello_2005b}, \citet{Starkenburg_2014}, and \citet{Hansen_2016b} showed that most CEMP-$s$ stars are in fact member of binary systems, confirming results that apply to stars over a wide range of metallicities \citep[e.g.,][]{Abate_2013, Lee_2014, Abate_2015c}. Some have argued that this mechanism is responsible for the carbon enhancement in {\it all} CEMP stars. \citet{Hansen_2016a} showed that Group II CEMP stars are not often associated with binary status. In contrast, \citet{Arentsen_2019} suggested that they may be wide (long-period) binaries, and thus difficult to detect with a limited number of radial-velocity measurements. \citet{Suda_2004} and \citet{Komiya_2007, Komiya_2020} have argued that mass transfer from a long-period binary companion could explain the origin of the Group II CEMP stars. The recent work by \citet{Aguado_2022} used carbon isotropic ratios to show that, for the mega metal-poor ([Fe/H] = $-6.2$)star SMSS 1605-1443, which we would classify as a Group III CEMP-no star, the carbon was {\it not} produced and transferred from an AGB binary companion.

There are also a variety of proposed mechanisms for producing carbon-enriched material for Group II stars in the early universe, including massive rapidly rotating stars and various SNe scenarios \citep{Umeda_2003, Umeda_2005, Chieffi_2004, Iwamoto_2005, Meynet_2006, NOMOTO_2006, Tominaga_2007, Heger_2010, Meynet_2010, Limongi_2012, Ishigaki_2014b, Tominaga_2014, Choplin_2016, Salvadori_2016, Kobayashi_2020}.   

If the mass-transfer scenario applies to both Groups I and II CEMP CDTGs, they should both exhibit large dispersions in their corrected carbon-abundance distributions, as they would be due to local, and not global pollution events. A change in the abundances of a single star (or even multiple stars) in a CDTG would not affect the rest of the stars, leading to a wide dispersion in the elemental abundances.  Based on our statistical analysis of the Group I and II CEMP dynamical clusters, this interpretation is unlikely, as they exhibit quite different behaviors. In particular, the small, and statistically significant, dispersion of the observed \textbf{corrected} carbon abundances within the Group II CDTGs points to the carbon coming from the environment that the stars formed in, rather than from enriched material transferred from binary companions. The dispersion in the mean [C/Fe$]_{c}$ for Group II CDTGs is already at the measurement limit for the observations, so it does not appear plausible to explain this low dispersion with anything but natal carbon abundances.

Note that the models described by \citet{Komiya_2020} could produce carbon enhancements at low metallicity without considering a significant mass transfer, if some of their uncertain modeling parameters are considered. As they point out, the Fe yield of faint (mixing and fallback) supernovae is highly affected by the ejection factor, which controls the fraction of materials ejected from the region between the mass cut and the outer boundary of the mixing region. Also, [C/Fe] ratios could be affected by the diffusion coefficient, the amount of mass swept up by SNe, and the nature of the IMF. Their optimum model, with lower swept-up mass and diffusion coefficient could reproduce the trend of [C/Fe], while this model overproduces stars with [Fe/H] $< -$4 (Figure 14 of \citealt{Komiya_2020}).

Taking into account these uncertainties, the contradiction between our results and the mass-transfer scenario might be mitigated. \citet{Komiya_2020} only studied the effect of mass transfer in models without faint SNe. If they include mass transfer in models that include faint SNe, they would not need to assume a high rate of binary mass transfer to explain CEMP-no stars. In this case, the mass transfer, even if it occurs, may not significantly affect the low observed dispersion of carbon in Group II CDTGs.

Our results suggest the hypothesis that Group II CEMP stars are formed from the gas enriched by rapidly rotating massive stars or the ejecta of SNe with large carbon enhancements and depleted Fe production. Faint (mixing and fallback) SNe are one of the astrophysical candidates that have been suggested to be responsible \citep{Umeda_2003, Umeda_2005, Iwamoto_2005, Tominaga_2007, Ishigaki_2014b}. These SNe eject a small amount of Fe due to the large amount of fallback, resulting in ejecta with large [C/Fe] abundance ratios. Indeed, \citet{Tominaga_2007} successfully reproduced the observed abundance pattern of the CEMP-no star CS~29498-043 \citep{Aoki_2004} with their 25 $M_{\sun}$ faint SN model. Thus, although the predicted [C/Fe] ratios are highly dependent on the choice of uncertain parameters, the ejecta from faint SNe could serve as one of the primary production sites for Group II CEMP stars.

Several studies have attempted to explain CEMP-no star formation solely from SNe ejecta \citep[e.g.,][]{Salvadori_2015, Sarmento_2016, Chiaki_2017, deBennassuti_2017, Sharma_2018, Sharma_2018b, Hartwig_2019, Sarmento_2019, Chiaki_2020}. Using the \textsc{eagle} cosmological hydrodynamical simulation \citep{Schaye_2015}, \citet{Sharma_2018} found that low-metallicity core-collapse SNe with low Fe yields contribute to forming CEMP-no stars. Stars formed from such SNe exhibited high [C/Fe] and low [C/O], similar to CEMP-no stars' observed properties. \citet{Hartwig_2019} proposed a new scenario for forming CEMP-no stars. Their analytical estimates have shown that such stars could be formed from the carbon-rich material before those formed from carbon-normal gas in the inhomogeneous interstellar medium, due to the shorter cooling time of the carbon-enhanced gas. In this manner, CEMP-no stars could be formed from carbon-normal SNe.

Our results clearly favor one or more of the rapidly rotating massive-star and SNe scenarios, rather than the binary mass-transfer scenario(s). Note that, as shown in Table~\ref{tab:binomial_probability_table_full}, the dispersion of [Mg/Fe] for the Group II CEMP stars is of marginal statistical significance, as reflected in the GEAD probability, unlike the highly significant behavior of [C/Fe]$_{c}$. While core-collapse SNe synthesize Mg, the lack of strong significance in [Mg/Fe] does not rule out this hypothesis for the stars in group Group II CDTGs, in part due to the limited number of Group II CDTGs (5) with measured [Mg/Fe] dispersions in our sample. Recent cosmological zoom-in simulations of MW-like galaxies show a low expected scatter of [Mg/Fe] in low-mass disrupted dwarf galaxies \citep{Hirai_2022}, which would make detecting the similarities difficult. 

Taking the available evidence as a whole, informed by the work in the current paper, we propose that most of the CEMP stars in Group II CDTGs (primarily CEMP-no stars) are formed from rapidly rotating, high-mass stars or the ejecta of various SNe, while stars in Group I CDTGs (primarily CEMP-$s$ stars) are formed from binary mass transfer from an evolved companion. To confirm and refine these ideas, future studies need to explore the chemical nature of CDTGs with other elements, and conduct more comprehensive galaxy-formation simulations with improved modeling of CEMP star formation.

\section{Conclusions}\label{sec:Conclusions}

In this work, we have assembled a sample of CEMP stars with elemental abundances for a select number of species derived from moderate- to high-resolution spectroscopy. We then derived space velocities for the sample to estimate each star's dynamical parameters ($E,J_{r},J_{\phi},J_{z}$) using \texttt{AGAMA}. Dynamical clustering of this sample using HDBSCAN resulted in the identification of 40 CDTGs. We found CDTGs to be associated with the following MW substructures: GSE, LMS-1, Thamnos, the Metal-Weak Thick Disk, the Splashed Disk and I'itoi. We only found a few CEMP CDTGs (4) to be associated with globular clusters, consistent with many previous claims of the lack of carbon-enhanced stars in these dense environments.

Finally, we separated the full sample of stars with available dynamical parameters into the CEMP morphological groups in the Yoon-Beers diagram of $A$(C)$_c$ vs. [Fe/H], and repeated the clustering exercise. Although the numbers of such clusters are reduced in total, the CEMP Group I CDTGs exhibit high [C/Fe]$_c$ dispersions, leading to the interpretation that the stars in these groups (primarily CEMP-$s$) likely arise from local enrichment events, such as mass transfer, in environments with extended star formation.  In stark contrast, the CEMP Group II CDTGs exhibit such small dispersions in [C/Fe]$_c$ that we argue they cannot have arisen from local events, but formed instead from natal gas enriched by high-mass early generation rapidly rotating stars or SNe explosions in environments that did not support extended star formation, such as low-mass satellite galaxies.

\vspace{2.0cm}

\begin{acknowledgements}

We thank an anonymous referee for comments that helped clarify our presentation. 
J.Z., T.C.B., D.S., D.G., Y.H., J.Y., M.M., C.P., T.P., and S.C. acknowledge partial support for this work from grant PHY 14-30152; Physics Frontier Center/JINA Center for the Evolution of the Elements (JINA-CEE), and OISE-1927130: The International Research Network for Nuclear Astrophysics (IReNA), awarded by the US National Science Foundation. Y.H. is partly supported by JSPS KAKENHI Grant Numbers JP21J00153, JP20K14532, JP21H04499, JP21K03614, and JP22H01259.
The work of V.M.P. is supported by NOIRLab, which is managed by the Association of Universities for Research in Astronomy (AURA) under a cooperative agreement with the National Science Foundation.

\end{acknowledgements}

\section{Appendix}


Here we present the tables for the Initial Sample of CEMP stars (Table~\ref{tab:initial_data_descript}) and the Final Sample of CEMP stars (Table~\ref{tab:final_data_descript}). 
In the print edition, only the table descriptions are provided; the full tables are available only in electronic form.


We show the results obtained for the Group I CDTGs 
(Table~\ref{tab:g1_cluster_results_stub}), their dynamical parameters 
(Table~\ref{tab:g1_cluster_orbit}), and the means, dispersions, and Inter-Quartile Ranges of the CDTGs (Table~\ref{tab:g1_cluster_mean}). Similar tables for the Group II CDTGs are provided in 
Table~\ref{tab:g2_cluster_results_stub}, Table~\ref{tab:g2_cluster_orbit}, and Table~\ref{tab:g2_cluster_mean}, respectively.

\include{Tables/initial_data_description_table}

\input{Tables/final_data_description_table}



\input{Tables/group1_cluster_stellar_results_stub_table}

\input{Tables/g1_cluster_orbital_table}


\input{Tables/g1_cluster_mean_table}

\input{Tables/group2_cluster_stellar_results_stub_table}

\input{Tables/g2_cluster_orbital_table}


\input{Tables/g2_cluster_mean_table}






\clearpage
\bibliography{references.bib}
\bibliographystyle{aasjournal}

\end{document}

%% file: Tables/group_stub.tex
\begin{deluxetable*}{l  l  c  c  c  c  c  c  c  c  c  c}
\tabletypesize{\scriptsize}
\tablecaption{Identified CEMP stars and their Group Associations \label{tab:group_stub}}
\tablehead{\colhead{Name} & \colhead{Group} & \colhead{T$_{\rm eff}$ (K)} & \colhead{log $g$} & \colhead{[Fe/H]} & \colhead{[C/Fe]$_{c}$} & 
\colhead{$A$(C)$_{c}$} & \colhead{[Mg/Fe]} & \colhead{[Sr/Fe]} & \colhead{[Y/Fe]} & \colhead{[Ba/Fe]} & \colhead{[Eu/Fe]}}
\startdata
HD 224959 & I & $5050$ & $2.100$ & $-2.42$ & $+2.04$ & $8.05$ & $+0.39$ & $+1.50$ & $\dots$ & $+2.07$ & $+2.01$ \\
SDSS J000219.87$+$292851.8 & I & $6150$ & $4.000$ & $-3.26$ & $+2.63$ & $7.80$ & $+0.36$ & $+0.27$ & $\dots$ & $+1.83$ & $\dots$ \\
BPS CS 29503$-$0010 & I & $6050$ & $3.660$ & $-1.70$ & $+1.65$ & $8.38$ & $\dots$ & $+1.13$ & $\dots$ & $+1.81$ & $+1.69$ \\
HE 0002$-$1037 & I & $4673$ & $1.280$ & $-3.75$ & $+3.43$ & $8.11$ & $\dots$ & $\dots$ & $\dots$ & $\dots$ & $\dots$ \\
BPS CS 31070$-$0073* & I/II & $6190$ & $3.860$ & $-2.55$ & $+1.34$ & $7.22$ & $+0.53$ & $+1.75$ & $\dots$ & $+2.28$ & $+2.74$ \\
HE 0007$-$1832 & I & $6500$ & $3.800$ & $-2.79$ & $+2.79$ & $8.43$ & $+0.55$ & $+0.08$ & $\dots$ & $+0.04$ & <+1.70 \\
HE 0010$-$3422 & I & $5400$ & $3.100$ & $-2.78$ & $+1.93$ & $7.58$ & $+0.34$ & $+0.77$ & $\dots$ & $+1.46$ & $+1.64$ \\
HE 0012$-$1441 & I & $5750$ & $3.500$ & $-2.52$ & $+1.71$ & $7.62$ & $+0.83$ & $\dots$ & $\dots$ & $+1.10$ & $\dots$ \\
HE 0013$-$0257 & II & $4500$ & $0.500$ & $-3.82$ & $+0.93$ & $5.54$ & $+0.68$ & $-0.46$ & $\dots$ & $-1.16$ & <+0.67 \\
HE 0015$+$0048 & II & $4600$ & $0.900$ & $-3.07$ & $+1.27$ & $6.63$ & $+0.65$ & $-1.07$ & $\dots$ & $-1.18$ & <+0.75 \\
HE 0017$-$4346 & I & $6198$ & $3.800$ & $-3.07$ & $+3.02$ & $8.38$ & $+0.86$ & $+0.84$ & $\dots$ & $+1.23$ & <+0.99 \\
HE 0017$+$0055 & I & $4261$ & $0.800$ & $-2.47$ & $+2.80$ & $8.76$ & $\dots$ & $\dots$ & $\dots$ & $+2.30$ & $+2.14$ \\
SMSS J002148.06$-$471132.1 & II & $4765$ & $1.400$ & $-3.17$ & $+0.78$ & $6.04$ & $+0.43$ & $+0.02$ & $\dots$ & $-1.18$ & <+0.50 \\
2MASS J00224486$-$1724290* & II & $4765$ & $1.550$ & $-4.05$ & $+2.07$ & $6.45$ & $+1.03$ & $-0.73$ & $\dots$ & $-1.10$ & $+0.25$ \\
SDSS J002314.00$+$030758.0* & III & $5997$ & $4.600$ & $-5.80$ & $+3.76$ & $6.39$ & $+3.33$ & $\dots$ & $\dots$ & $\dots$ & $\dots$ \\
HE 0024$-$2523 & I & $6625$ & $4.300$ & $-2.72$ & $+2.69$ & $8.40$ & $+0.71$ & $+0.44$ & $\dots$ & $+1.41$ & <+0.07 \\
BPS CS 22882$-$0012 & I/II & $6290$ & $3.800$ & $-2.75$ & $+1.10$ & $6.78$ & $+0.36$ & $+0.56$ & $\dots$ & $+0.61$ & <+1.23 \\
BPS CS 31062$-$0050 & I & $5600$ & $3.000$ & $-2.32$ & $+2.12$ & $8.23$ & $+0.59$ & $+0.96$ & $\dots$ & $+2.35$ & $+1.87$ \\
2MASS J00305267$-$1007042 & I/II & $4831$ & $1.480$ & $-2.35$ & $+0.88$ & $6.96$ & $+0.24$ & $+0.50$ & $\dots$ & $-0.71$ & $+0.00$ \\
SDSS J003602.17$-$104336.2 & I & $6500$ & $4.500$ & $-2.50$ & $+2.37$ & $8.30$ & $+0.27$ & $-0.13$ & $\dots$ & $+0.37$ & $\dots$ \\
RAVE J003946.9$-$684957 & II & $4631$ & $1.160$ & $-2.62$ & $+0.73$ & $6.54$ & $\dots$ & $+0.43$ & $\dots$ & $-0.10$ & $-0.38$ \\
BPS CS 29497$-$0030 & I & $7000$ & $4.000$ & $-2.52$ & $+2.37$ & $8.28$ & $+0.35$ & $+1.30$ & $\dots$ & $+2.75$ & $+1.71$ \\
BPS CS 29497$-$0034 & I & $4800$ & $1.800$ & $-2.90$ & $+2.78$ & $8.31$ & $+0.70$ & $+1.05$ & $\dots$ & $+1.98$ & $+1.79$ \\
G 270$-$51 & I & $6000$ & $3.800$ & $-2.74$ & $+2.39$ & $8.08$ & $+0.43$ & $+0.18$ & $\dots$ & $+1.96$ & $+1.39$ \\
2MASS J00453930$-$7457294 & I & $4947$ & $2.010$ & $-2.00$ & $+0.99$ & $7.42$ & $\dots$ & $+0.83$ & $\dots$ & $+0.37$ & $+0.55$
\enddata
\tablecomments{This table is a stub; the full table is available in the electronic edition. Stars with names that end with * are not in the Final Sample.}
\end{deluxetable*}

%% file: Tables/cluster_summary_table.tex
\startlongtable
\begin{deluxetable*}{l  r  r  c}
\tabletypesize{\scriptsize}
\tablecaption{Identified CDTGs \label{tab:cluster_summary}}
\tablehead{\colhead{CDTG} & \colhead{$N$ Stars} & \colhead{Confidence} & \colhead{Associations}}
\startdata
1 & $17$ & $100.0\%$ & H22:DTC-24, IR18:A, DG21:CDTG-11, H22:DTC-3, DS22b:DTG-51, DG21:CDTG-16\\
& & &  IR18:C, H22:DTC-4, DG21:CDTG-3\\
2 & $12$ & $99.8\%$ & GL21:DTG-31, H22:DTC-16, DS22a:DTG-50, GC21:Sequoia, KM22:Fimbulthul, DS22b:DTG-18\\
3 & $11$ & $58.2\%$ & I'itoi, GM18b:Rg5, GL21:DTG-6, KM22:C-3, KM22:Gaia-6\\
4 & $10$ & $91.1\%$ & GSE, AH17:VelHel-6, H22:DTC-3, KM22:C-3, DS22b:DTG-137\\
5 & $10$ & $78.5\%$ & GSE, KM22:C-3\\
6 & $10$ & $99.2\%$ & Thamnos, DG21:CDTG-27, ZY20b:DTG-33, HL20:GR-2, H22:DTC-16, GC21:Sequoia, KM22:Gunnthra\\
& & &  KM22:C-3, KM22:Gaia-6, EV21:NGC~6235, EV21:NGC~6356\\
7 & $9$ & $83.5\%$ & GSE, SL22:3, DS22b:DTG-27, DS22b:DTG-7\\
8 & $9$ & $98.3\%$ & GSE, GL21:DTG-34, H22:DTC-2, DS22b:DTG-95, DG21:CDTG-28, DS22b:DTG-30\\
& & &  GC21:Sausage, DG21:CDTG-9, DG21:CDTG-13, GL21:DTG-28, SL22:1\\
9 & $9$ & $99.9\%$ & LMS-1, GL21:DTG-17, DS22b:DTG-38, FS19:IH, HL19:GL-1, KM22:C-3\\
10 & $8$ & $100.0\%$ & DS22b:DTG-22\\
11 & $8$ & $68.2\%$ & GSE, H22:DTC-15, DS22b:DTG-79, GM17:Comoving, DS22a:DTG-29, DS22b:DTG-20, DS22b:DTG-135\\
12 & $7$ & $78.5\%$ & HL19:GL-1, KM22:C-3, KM22:NGC7089\\
13 & $7$ & $100.0\%$ & DS22b:DTG-35\\
14 & $7$ & $82.8\%$ & DS22a:DTG-6, GL21:DTG-4, H22:DTC-1, KM22:C-3\\
15 & $6$ & $64.7\%$ & LMS-1, H22:DTC-29, FS19:OH, KM22:C-3\\
16 & $6$ & $80.6\%$ & GSE, DG21:CDTG-21, H22:DTC-3, KM22:NGC7089\\
17 & $6$ & $93.1\%$ & GSE, SL22:59, H22:DTC-22, SM20:Sausage, GC21:Sausage\\
18 & $6$ & $98.7\%$ & DS22a:DTG-6\\
19 & $5$ & $100.0\%$ & new\\
20 & $5$ & $54.8\%$ & GSE, DS22b:DTG-12, EV21:Ryu~879~(RLGC~2)\\
21 & $5$ & $99.7\%$ & Thamnos, DS22b:DTG-170, DS22a:DTG-51, GL21:DTG-31, SM20:SeqG1, GC21:Sequoia\\
22 & $5$ & $77.2\%$ & Splashed Disk\\
23 & $5$ & $68.9\%$ & FS19:IH, KM22:C-3\\
24 & $5$ & $100.0\%$ & DS22b:DTG-161\\
25 & $4$ & $59.1\%$ & GSE, GL21:DTG-11, SM20:Sausage, DG21:CDTG-21, GL21:DTG-21, DS22b:DTG-98, SL22:3\\
26 & $4$ & $69.4\%$ & DS22b:DTG-4, ZY20b:DTG-44, GM17:Comoving, KM22:C-3, DS22a:DTG-4\\
27 & $4$ & $94.9\%$ & new\\
28 & $4$ & $100.0\%$ & GSE\\
29 & $4$ & $60.9\%$ & GSE, GL21:DTG-34, GL21:DTG-35, FS19:P, SL22:3, GC21:Sausage, DS22b:DTG-168\\
30 & $4$ & $68.2\%$ & GSE, FS19:OH, FS19:IH, GM17:Comoving, SM20:Sausage\\
31 & $4$ & $40.3\%$ & GSE, DS22a:DTG-8, DS22b:DTG-156, GC21:Sausage, KM22:NGC7089\\
32 & $4$ & $62.5\%$ & FS19:IH, KM22:NGC7089, EV21:FSR~1758\\
33 & $4$ & $71.2\%$ & GSE, GL21:DTG-20\\
34 & $3$ & $35.5\%$ & GSE, GM17:Comoving, SM20:Sausage, GC21:Sausage, GL21:DTG-11, KM22:Hrid, DS22a:DTG-27\\
& & &  DS22b:DTG-50, SL22:7\\
35 & $3$ & $61.6\%$ & GSE, GL21:DTG-38, ZY20b:DTG-40, GC21:Sausage, DS22b:DTG-71\\
36 & $3$ & $51.7\%$ & GSE, DS22b:DTG-27, DS22b:DTG-45, GC21:Sausage, DS22a:DTG-17\\
37 & $3$ & $68.4\%$ & H22:DTC-4\\
38 & $3$ & $57.7\%$ & MWTD, DS22b:DTG-43, DS22a:DTG-44\\
39 & $3$ & $58.6\%$ & GSE, KM22:NGC7089\\
40 & $3$ & $50.5\%$ & GL21:DTG-19, GC21:Sequoia, EV21:Pfleiderer~2\\
\enddata
\tablecomments{We adopt the nomenclature for previously identified DTGs and CDTGs from \cite{Yuan_2020b}.}\end{deluxetable*}

%% file: Tables/cluster_stellar_results_stub_table.tex
\begin{deluxetable*}{l  c  c  c  c  c  c  c  c}
\tablecaption{CDTGs Identified by \texttt{HDBSCAN} \label{tab:cluster_results_stub}}
\tablehead{\colhead{NAME} & \colhead{[Fe/H]} & \colhead{[C/Fe]$_{c}$} & \colhead{[Mg/Fe]} & \colhead{[Sr/Fe]} & \colhead{[Y/Fe]} & \colhead{[Ba/Fe]} & \colhead{[Eu/Fe]}}
\startdata
\multicolumn{8}{c}{CDTG-$1$} \\
\multicolumn{8}{c}{Structure: Unassigned Structure} \\
\multicolumn{8}{c}{Group Assoc: CDTG-3: \citet{Gudin_2021}} \\
\multicolumn{8}{c}{Group Assoc: DTG-51: \citet{Shank_2022b}} \\
\multicolumn{8}{c}{Stellar Assoc: BPS CS 31078-0018 (DTC-24: \citealt{Hattori_2022})} \\
\multicolumn{8}{c}{Stellar Assoc: HE 0430-4901 (A: \citealt{Roederer_2018})} \\
\multicolumn{8}{c}{Stellar Assoc: HE 0430-4901 (CDTG-11: \citealt{Gudin_2021})} \\
\multicolumn{8}{c}{Stellar Assoc: HE 0430-4901 (DTC-3: \citealt{Hattori_2022})} \\
\multicolumn{8}{c}{Stellar Assoc: J100824.90-231412.0 (DTG-51: \citealt{Shank_2022b})} \\
\multicolumn{8}{c}{Stellar Assoc: RAVE J192819.9-633935* (CDTG-16: \citealt{Gudin_2021})} \\
\multicolumn{8}{c}{Stellar Assoc: CS 22945-017 (C: \citealt{Roederer_2018})} \\
\multicolumn{8}{c}{Stellar Assoc: BPS CS 22945-0017 (CDTG-16: \citealt{Gudin_2021})} \\
\multicolumn{8}{c}{Stellar Assoc: BPS CS 22945-0017 (DTC-4: \citealt{Hattori_2022}) } \\
\multicolumn{8}{c}{Globular Assoc: No Globular Associations} \\
\multicolumn{8}{c}{Dwarf Galaxy Assoc: No Dwarf Galaxy Associations} \\
HE 0017$-$4346 & $-3.07$ & $+3.02$ & $+0.86$ & $+0.84$ & $\dots$ & $+1.23$ & $+0.99$ \\
RAVE J014908.0$-$491143 & $-2.94$ & $+0.77$ & $\dots$ & $-0.45$ & $\dots$ & $-0.59$ & $+0.09$ \\
BPS CS 31078$-$0018 & $-3.02$ & $+0.74$ & $\dots$ & $\dots$ & $\dots$ & $+0.08$ & $\dots$ \\
HE 0430$-$4901 & $-3.10$ & $+1.00$ & $\dots$ & $\dots$ & $\dots$ & $\dots$ & $\dots$ \\
RAVE J053817.0$-$751621 & $-2.03$ & $+0.78$ & $\dots$ & $-0.63$ & $\dots$ & $-0.24$ & $+0.44$ \\
LAMOST J070542.30$+$255226.6 & $-3.19$ & $+1.78$ & $+1.04$ & $\dots$ & $\dots$ & $\dots$ & $\dots$ \\
2MASS J09294972$-$2905589 & $-2.32$ & $+0.75$ & $+0.23$ & $-0.36$ & $\dots$ & $-0.37$ & $+0.14$ \\
RAVE J100824.9$-$231412 & $-1.95$ & $+0.78$ & $\dots$ & $\dots$ & $\dots$ & $+0.36$ & $-0.19$ \\
HE 1105$+$0027 & $-2.42$ & $+1.96$ & $+0.45$ & $+0.83$ & $\dots$ & $+2.40$ & $+1.80$ \\
2MASS J11580127$-$1522179 & $-2.41$ & $+1.01$ & $+0.20$ & $-0.37$ & $\dots$ & $-1.07$ & $+0.15$ \\
GALAH 150209004501153 & $-1.61$ & $+0.79$ & $\dots$ & $\dots$ & $+1.54$ & $+0.36$ & $\dots$ \\
Pristine J209.9364$+$15.9251 & $-2.25$ & $+2.18$ & $+0.22$ & $+0.62$ & $\dots$ & $\dots$ & $\dots$ \\
HE 1413$-$1954 & $-3.22$ & $+1.44$ & $\dots$ & $-0.34$ & $\dots$ & $\dots$ & $\dots$ \\
HD 187216 & $-2.50$ & $+1.70$ & $\dots$ & $\dots$ & $\dots$ & $+2.00$ & $+1.66$ \\
RAVE J192819.9$-$633935 & $-2.23$ & $+0.76$ & $+0.61$ & $\dots$ & $\dots$ & $+0.09$ & $+0.45$ \\
HE 2150$-$0825 & $-1.98$ & $+1.31$ & $+0.33$ & $+0.75$ & $\dots$ & $+1.65$ & $\dots$ \\
BPS CS 22945$-$0017 & $-2.73$ & $+1.78$ & $+0.26$ & $+0.38$ & $\dots$ & $+0.48$ & $+1.13$ \\
$\boldsymbol{\mu \pm \sigma ([X/Y])}$ & $\boldsymbol{-2.52\pm0.51}$ & $\boldsymbol{+1.09\pm0.62}$ & $\boldsymbol{+0.34\pm0.28}$ & $\boldsymbol{+0.11\pm0.62}$ & $\boldsymbol{+1.54\pm \dots}$ & $\boldsymbol{+0.41\pm1.04}$ & $\boldsymbol{+0.57\pm0.71}$\\
\pagebreak
\enddata
\tablecomments{$\boldsymbol{\mu}$ and $\boldsymbol{\sigma}$ represent the biweight estimates of the location and scale for the abundances in the CDTG.}\tablecomments{This table is a stub; the full table is available in the electronic edition.}\end{deluxetable*}

%% file: Tables/cluster_orbital_table.tex
\startlongtable
\begin{deluxetable*}{l  r  c  c  r  r}
\tablecaption{Cluster Dynamical Parameters Determined by \texttt{AGAMA} \label{tab:cluster_orbit}}
\tablehead{\colhead{Cluster} & \colhead{$N$ Stars} & \colhead{($\langle$v$_{\text{r}}\rangle$,$\langle$v$_{\phi}\rangle$,$\langle$v$_{\text{z}}\rangle$)} & \colhead{($\langle$J$_{\text{r}}\rangle$,$\langle$J$_{\phi}\rangle$,$\langle$J$_{\text{z}}\rangle$)} & \colhead{$\langle$E$\rangle$} & \colhead{$\langle$ecc$\rangle$}\\
\colhead{} & \colhead{} & \colhead{($\sigma_{\langle\text{v}_{\text{r}}\rangle}$,$\sigma_{\langle\text{v}_{\phi}\rangle}$,$\sigma_{\langle\text{v}_{\text{z}}\rangle}$)} & \colhead{($\sigma_{\langle\text{J}_{\text{r}}\rangle}$,$\sigma_{\langle\text{J}_{\phi}\rangle}$,$\sigma_{\langle\text{J}_{\text{z}}\rangle}$)} & \colhead{$\sigma_{\langle\text{E}\rangle}$} & \colhead{$\sigma_{\langle\text{ecc}\rangle}$}\\
\colhead{} & \colhead{} & \colhead{(km s$^{-1}$)} & \colhead{(kpc km s$^{-1}$)} & \colhead{(10$^{5}$ km$^{2}$ s$^{-2}$)} & \colhead{}}
\startdata
CDTG$-1$ & $17$ & ($56.3$,$154.2$,$-11.2$) & ($161.2$,$1273.1$,$63.0$) & $-1.642$ & $0.38$ \\
 & & ($74.3$,$18.2$,$50.3$) & ($61.8$,$69.6$,$40.1$) & $0.024$ & $0.07$ \\
CDTG$-2$ & $12$ & ($-18.5$,$-88.6$,$-19.1$) & ($291.4$,$-562.3$,$83.2$) & $-1.824$ & $0.63$ \\
 & & ($84.5$,$25.3$,$40.9$) & ($80.3$,$87.1$,$41.9$) & $0.040$ & $0.07$ \\
CDTG$-3$ & $11$ & ($-15.6$,$-134.2$,$-95.6$) & ($156.7$,$-911.0$,$707.2$) & $-1.534$ & $0.37$ \\
 & & ($104.5$,$48.2$,$170.6$) & ($79.4$,$178.2$,$128.5$) & $0.041$ & $0.11$ \\
CDTG$-4$ & $10$ & ($-60.3$,$69.3$,$-62.1$) & ($396.6$,$550.6$,$165.3$) & $-1.722$ & $0.70$ \\
 & & ($91.0$,$21.5$,$75.7$) & ($27.5$,$93.6$,$44.6$) & $0.021$ & $0.03$ \\
CDTG$-5$ & $10$ & ($-63.4$,$-13.5$,$67.3$) & ($314.9$,$-63.4$,$628.9$) & $-1.728$ & $0.72$ \\
 & & ($93.0$,$22.3$,$139.7$) & ($42.5$,$146.9$,$53.5$) & $0.024$ & $0.06$ \\
CDTG$-6$ & $10$ & ($1.4$,$-120.7$,$-13.2$) & ($137.0$,$-875.8$,$266.7$) & $-1.704$ & $0.41$ \\
 & & ($55.7$,$26.5$,$119.5$) & ($69.3$,$140.7$,$99.1$) & $0.049$ & $0.10$ \\
CDTG$-7$ & $9$ & ($2.2$,$33.6$,$26.5$) & ($499.3$,$259.5$,$74.2$) & $-1.814$ & $0.84$ \\
 & & ($61.2$,$15.2$,$78.2$) & ($71.1$,$119.0$,$29.5$) & $0.018$ & $0.08$ \\
CDTG$-8$ & $9$ & ($148.2$,$-22.6$,$-13.2$) & ($811.8$,$-186.6$,$20.3$) & $-1.678$ & $0.92$ \\
 & & ($35.4$,$10.4$,$37.3$) & ($51.7$,$88.7$,$63.9$) & $0.036$ & $0.04$ \\
CDTG$-9$ & $9$ & ($60.4$,$100.0$,$-68.5$) & ($180.3$,$699.4$,$584.0$) & $-1.620$ & $0.44$ \\
 & & ($56.5$,$33.0$,$174.3$) & ($45.3$,$55.8$,$90.2$) & $0.038$ & $0.06$ \\
CDTG$-10$ & $8$ & ($53.3$,$111.9$,$3.6$) & ($242.0$,$879.4$,$144.2$) & $-1.707$ & $0.52$ \\
 & & ($55.4$,$7.3$,$64.6$) & ($8.3$,$47.9$,$29.2$) & $0.022$ & $0.01$ \\
CDTG$-11$ & $8$ & ($8.4$,$27.1$,$10.8$) & ($400.9$,$146.8$,$183.3$) & $-1.861$ & $0.88$ \\
 & & ($133.7$,$18.4$,$89.6$) & ($58.7$,$84.3$,$33.8$) & $0.050$ & $0.06$ \\
CDTG$-12$ & $7$ & ($-37.2$,$83.2$,$-16.2$) & ($268.9$,$604.6$,$302.0$) & $-1.715$ & $0.58$ \\
 & & ($90.8$,$8.7$,$81.3$) & ($40.1$,$60.3$,$52.2$) & $0.018$ & $0.03$ \\
CDTG$-13$ & $7$ & ($-15.9$,$188.7$,$-2.2$) & ($31.8$,$1552.4$,$37.3$) & $-1.634$ & $0.17$ \\
 & & ($18.5$,$2.5$,$37.0$) & ($7.5$,$61.9$,$28.3$) & $0.024$ & $0.03$ \\
CDTG$-14$ & $7$ & ($-2.8$,$15.9$,$-218.0$) & ($246.5$,$-7.9$,$1410.4$) & $-1.454$ & $0.44$ \\
 & & ($123.0$,$89.4$,$76.2$) & ($87.1$,$440.2$,$105.3$) & $0.055$ & $0.06$ \\
CDTG$-15$ & $6$ & ($-67.5$,$73.2$,$-95.7$) & ($455.8$,$413.2$,$972.8$) & $-1.487$ & $0.60$ \\
 & & ($123.5$,$54.2$,$131.9$) & ($113.3$,$176.5$,$49.1$) & $0.051$ & $0.10$ \\
CDTG$-16$ & $6$ & ($-31.1$,$51.3$,$23.1$) & ($810.1$,$457.0$,$188.6$) & $-1.530$ & $0.83$ \\
 & & ($174.9$,$15.7$,$78.6$) & ($16.7$,$63.8$,$59.7$) & $0.035$ & $0.03$ \\
CDTG$-17$ & $6$ & ($-155.2$,$-21.4$,$58.9$) & ($1284.8$,$-158.1$,$515.3$) & $-1.368$ & $0.95$ \\
 & & ($153.1$,$8.7$,$144.5$) & ($72.6$,$80.5$,$97.6$) & $0.028$ & $0.03$ \\
CDTG$-18$ & $6$ & ($87.8$,$-12.0$,$165.9$) & ($148.8$,$-90.4$,$1917.9$) & $-1.407$ & $0.30$ \\
 & & ($113.5$,$20.8$,$214.4$) & ($163.8$,$187.0$,$113.4$) & $0.049$ & $0.14$ \\
CDTG$-19$ & $5$ & ($17.5$,$178.6$,$41.8$) & ($42.1$,$1122.0$,$197.8$) & $-1.703$ & $0.20$ \\
 & & ($45.7$,$34.7$,$112.3$) & ($33.6$,$21.2$,$4.2$) & $0.020$ & $0.09$ \\
CDTG$-20$ & $5$ & ($51.6$,$29.9$,$3.2$) & ($618.5$,$279.8$,$194.6$) & $-1.682$ & $0.86$ \\
 & & ($56.3$,$12.5$,$78.7$) & ($21.6$,$138.9$,$16.6$) & $0.033$ & $0.06$ \\
CDTG$-21$ & $5$ & ($-18.4$,$-128.9$,$28.0$) & ($254.5$,$-901.7$,$80.4$) & $-1.688$ & $0.54$ \\
 & & ($96.6$,$15.6$,$66.9$) & ($28.0$,$79.0$,$43.4$) & $0.033$ & $0.00$ \\
CDTG$-22$ & $5$ & ($-41.8$,$74.5$,$95.7$) & ($76.0$,$287.0$,$491.7$) & $-1.873$ & $0.44$ \\
 & & ($21.6$,$29.1$,$29.9$) & ($33.2$,$45.6$,$70.7$) & $0.058$ & $0.07$ \\
CDTG$-23$ & $5$ & ($38.0$,$85.6$,$60.9$) & ($77.4$,$420.3$,$672.2$) & $-1.739$ & $0.34$ \\
 & & ($98.2$,$3.3$,$144.4$) & ($68.7$,$45.7$,$61.4$) & $0.017$ & $0.20$ \\
CDTG$-24$ & $5$ & ($17.8$,$232.0$,$0.9$) & ($59.1$,$1995.8$,$57.5$) & $-1.503$ & $0.20$ \\
 & & ($63.5$,$19.8$,$54.5$) & ($7.3$,$144.6$,$44.5$) & $0.015$ & $0.02$ \\
CDTG$-25$ & $4$ & ($-12.9$,$-8.6$,$40.5$) & ($1279.9$,$-127.7$,$98.7$) & $-1.454$ & $0.92$ \\
 & & ($181.4$,$17.7$,$44.1$) & ($80.3$,$214.4$,$21.3$) & $0.041$ & $0.02$ \\
CDTG$-26$ & $4$ & ($9.9$,$-22.5$,$4.1$) & ($113.7$,$-164.1$,$983.4$) & $-1.672$ & $0.38$ \\
 & & ($117.7$,$28.2$,$176.3$) & ($50.9$,$207.4$,$27.6$) & $0.039$ & $0.08$ \\
CDTG$-27$ & $4$ & ($43.5$,$282.5$,$-3.3$) & ($224.6$,$2492.1$,$54.7$) & $-1.340$ & $0.34$ \\
 & & ($49.8$,$38.6$,$38.8$) & ($54.5$,$100.3$,$33.6$) & $0.034$ & $0.04$ \\
CDTG$-28$ & $4$ & ($-124.6$,$-22.5$,$-23.8$) & ($145.5$,$-57.3$,$113.6$) & $-2.174$ & $0.84$ \\
 & & ($24.1$,$31.3$,$81.8$) & ($61.2$,$98.6$,$4.1$) & $0.026$ & $0.02$ \\
CDTG$-29$ & $4$ & ($-118.0$,$12.5$,$-46.0$) & ($711.6$,$95.5$,$55.7$) & $-1.739$ & $0.93$ \\
 & & ($34.3$,$13.2$,$27.8$) & ($19.6$,$110.1$,$14.3$) & $0.011$ & $0.05$ \\
CDTG$-30$ & $4$ & ($-10.9$,$-11.4$,$58.0$) & ($1167.6$,$-104.5$,$801.2$) & $-1.338$ & $0.89$ \\
 & & ($233.9$,$12.6$,$160.9$) & ($87.7$,$128.6$,$35.8$) & $0.023$ & $0.03$ \\
CDTG$-31$ & $4$ & ($160.9$,$21.2$,$-13.7$) & ($978.3$,$164.9$,$314.1$) & $-1.503$ & $0.94$ \\
 & & ($47.6$,$2.9$,$122.9$) & ($59.5$,$53.2$,$43.3$) & $0.048$ & $0.02$ \\
CDTG$-32$ & $4$ & ($62.7$,$124.8$,$-34.3$) & ($519.3$,$1062.6$,$255.6$) & $-1.509$ & $0.61$ \\
 & & ($142.4$,$43.6$,$94.7$) & ($104.0$,$82.1$,$21.4$) & $0.031$ & $0.05$ \\
CDTG$-33$ & $4$ & ($35.7$,$-67.7$,$21.4$) & ($889.2$,$-573.3$,$91.3$) & $-1.503$ & $0.83$ \\
 & & ($219.1$,$17.6$,$67.2$) & ($20.8$,$148.6$,$31.4$) & $0.035$ & $0.04$ \\
CDTG$-34$ & $3$ & ($129.6$,$26.3$,$-70.7$) & ($1593.8$,$181.2$,$63.7$) & $-1.345$ & $0.93$ \\
 & & ($277.3$,$44.7$,$19.3$) & ($67.9$,$393.4$,$16.2$) & $0.036$ & $0.04$ \\
CDTG$-35$ & $3$ & ($69.8$,$-2.8$,$-17.6$) & ($663.7$,$-22.2$,$37.7$) & $-1.814$ & $0.98$ \\
 & & ($39.7$,$3.9$,$25.7$) & ($22.1$,$31.4$,$13.4$) & $0.014$ & $0.01$ \\
CDTG$-36$ & $3$ & ($19.5$,$12.3$,$8.0$) & ($516.0$,$81.5$,$103.8$) & $-1.851$ & $0.95$ \\
 & & ($97.3$,$8.8$,$53.2$) & ($34.3$,$59.3$,$23.3$) & $0.021$ & $0.03$ \\
CDTG$-37$ & $3$ & ($5.9$,$189.1$,$32.9$) & ($38.6$,$891.0$,$181.2$) & $-1.796$ & $0.23$ \\
 & & ($51.9$,$17.9$,$78.0$) & ($21.9$,$59.6$,$22.5$) & $0.023$ & $0.07$ \\
CDTG$-38$ & $3$ & ($32.0$,$111.4$,$41.5$) & ($174.9$,$701.8$,$63.1$) & $-1.861$ & $0.49$ \\
 & & ($27.3$,$8.8$,$26.3$) & ($18.5$,$61.3$,$5.5$) & $0.024$ & $0.04$ \\
CDTG$-39$ & $3$ & ($-222.0$,$88.6$,$-25.1$) & ($860.9$,$548.3$,$377.3$) & $-1.456$ & $0.80$ \\
 & & ($64.6$,$30.7$,$95.3$) & ($38.1$,$36.5$,$42.0$) & $0.016$ & $0.01$ \\
CDTG$-40$ & $3$ & ($61.2$,$-116.4$,$55.7$) & ($389.1$,$-1050.4$,$96.5$) & $-1.599$ & $0.58$ \\
 & & ($105.2$,$19.8$,$23.5$) & ($20.4$,$95.9$,$50.0$) & $0.014$ & $0.03$ \\
\enddata
\end{deluxetable*}

%% file: Tables/substructure_table.tex
\begin{deluxetable*}{l  r  r  r  r  r  r  r  r  c  c  c  r}
\tabletypesize{\tiny}
\tablecaption{Identified Milky Way Substructures \label{tab:substructures}}
\tablehead{\colhead{Substructure} & \colhead{$N$ Stars} & \colhead{$\langle$[Fe/H]$\rangle$} & \colhead{$\langle$[C/Fe]$_{\textit{c}}\rangle$} & \colhead{$\langle$[Mg/Fe]$\rangle$} & \colhead{$\langle$[Sr/Fe]$\rangle$} & \colhead{$\langle$[Y/Fe]$\rangle$} & \colhead{$\langle$[Ba/Fe]$\rangle$} & \colhead{$\langle$[Eu/Fe]$\rangle$} & \colhead{($\langle$v$_{\text{r}}\rangle$,$\langle$v$_{\phi}\rangle$,$\langle$v$_{\text{z}}\rangle$)} & \colhead{($\langle$J$_{\text{r}}\rangle$,$\langle$J$_{\phi}\rangle$,$\langle$J$_{\text{z}}\rangle$)} & \colhead{$\langle$E$\rangle$} & \colhead{$\langle$ecc$\rangle$}\\
\colhead{} & \colhead{} & \colhead{} & \colhead{} & \colhead{} & \colhead{} & \colhead{} & \colhead{} & \colhead{} & \colhead{($\sigma_{\langle\text{v}_{\text{r}}\rangle}$,$\sigma_{\langle\text{v}_{\phi}\rangle}$,$\sigma_{\langle\text{v}_{\text{z}}\rangle}$)} & \colhead{($\sigma_{\langle\text{J}_{\text{r}}\rangle}$,$\sigma_{\langle\text{J}_{\phi}\rangle}$,$\sigma_{\langle\text{J}_{\text{z}}\rangle}$)} & \colhead{$\sigma_{\langle\text{E}\rangle}$} & \colhead{$\sigma_{\langle\text{ecc}\rangle}$}\\
\colhead{} & \colhead{} & \colhead{} & \colhead{} & \colhead{} & \colhead{} & \colhead{} & \colhead{} & \colhead{} & \colhead{(km s$^{-1}$)} & \colhead{(kpc km s$^{-1}$)} & \colhead{(10$^{5}$ km$^{2}$ s$^{-2}$)} & \colhead{}}
\startdata
GSE & 99 & $-2.52$ & $+1.31$ & $+0.40$ & $+0.29$ & $+0.07$ & $+0.38$ & $+0.64$ & ($-11.6$,$17.9$,$2.6$) & ($701.8$,$104.8$,$233.4$) & $-1.666$ & $0.85$\\
 &  & $0.55$ & $0.73$ & $0.22$ & $0.74$ & $0.22$ & \textcolor{white}{+}$1.00$ & $0.62$ & ($152.1$,$48.1$,$89.3$) & ($351.2$,$305.5$,$217.2$) & $0.195$ & $0.10$\\
\hline
Thamnos & 15 & $-2.35$ & $+1.27$ & $+0.30$ & $-0.10$ & $NaN$ & $+0.62$ & $+0.68$ & ($-15.0$,$-124.0$,$-0.2$) & ($177.9$,$-915.3$,$201.7$) & $-1.702$ & $0.43$\\
 &  & $0.30$ & $0.53$ & $0.28$ & $1.33$ & $NaN$ & \textcolor{white}{+}$1.11$ & $0.54$ & ($69.3$,$23.0$,$102.0$) & ($75.9$,$146.9$,$117.9$) & $0.043$ & $0.10$\\
\hline
LMS-1 & 15 & $-3.11$ & $+1.44$ & $+0.54$ & $-0.27$ & $NaN$ & $-0.05$ & $+0.41$ & ($9.1$,$99.1$,$-50.5$) & ($293.2$,$579.4$,$742.5$) & $-1.567$ & $0.52$\\
 &  & $0.78$ & $0.63$ & $0.30$ & $1.05$ & $NaN$ & \textcolor{white}{+}$1.35$ & $0.57$ & ($106.7$,$47.7$,$137.7$) & ($152.1$,$185.2$,$201.7$) & $0.077$ & $0.11$\\
\hline
I'itoi & 11 & $-2.71$ & $+1.57$ & $+0.41$ & $+0.18$ & $+0.20$ & $+0.53$ & $+0.58$ & ($-21.7$,$-136.7$,$-37.7$) & ($184.5$,$-901.1$,$711.1$) & $-1.531$ & $0.38$\\
 &  & $0.33$ & $0.74$ & $0.06$ & $0.35$ & $0.00$ & \textcolor{white}{+}$1.35$ & $1.00$ & ($93.8$,$45.0$,$148.4$) & ($98.2$,$172.7$,$116.5$) & $0.040$ & $0.11$\\
\hline
Splashed Disk & 5 & $-2.69$ & $+1.23$ & $+1.17$ & $-0.18$ & $+0.09$ & $-0.03$ & $+0.88$ & ($-22.8$,$81.7$,$62.3$) & ($85.0$,$312.2$,$463.9$) & $-1.873$ & $0.45$\\
 &  & $0.63$ & $0.90$ & $0.59$ & $0.26$ & $0.34$ & \textcolor{white}{+}$0.98$ & $0.39$ & ($39.0$,$23.7$,$116.6$) & ($27.3$,$57.5$,$54.5$) & $0.055$ & $0.07$\\
\hline
MWTD & 3 & $-2.28$ & $+1.35$ & $+0.36$ & $+0.33$ & $NaN$ & $+0.33$ & $+0.51$ & ($32.0$,$111.4$,$41.5$) & ($174.9$,$701.8$,$63.1$) & $-1.861$ & $0.49$\\
 &  & $0.20$ & $0.79$ & $0.01$ & $0.27$ & $NaN$ & \textcolor{white}{+}$0.68$ & $0.27$ & ($27.3$,$8.8$,$26.3$) & ($18.5$,$61.3$,$5.5$) & $0.024$ & $0.04$\\
\hline
\enddata
\end{deluxetable*}

%% file: Tables/interesting_substructure_table.tex
\begin{deluxetable*}{l  l  l  c}
\tablecaption{Associations of Identified CDTGs \label{tab:interesting_substructure}}
\tablehead{\colhead{Structure} & \colhead{Reference} & \colhead{Associations} & \colhead{Identified CDTGs}}
\startdata
\multirow{6}{*}{MW Substructure} & \multirow{6}{*}{\citet{Naidu_2020}} & GSE & 4, 5, 7, 8, 11, 16, 17, 20, 25, 28, 29, 30, 31, 33, 34, 35, 36, 39\\ \cline{3-4}
 &  & LMS-1 & 9, 15\\ \cline{3-4}
 &  & Thamnos & 6, 21\\ \cline{3-4}
 &  & I'itoi & 3\\ \cline{3-4}
 &  & MWTD & 38\\ \cline{3-4}
 &  & Splashed Disk & 22\\ \cline{1-4}
\enddata
\end{deluxetable*}

%% file: Tables/interesting_substructure_groups_table.tex
\startlongtable
\begin{deluxetable*}{l  l  c}
\tablecaption{Associations of Identified CDTGs with Previous Groups \label{tab:interesting_groups_substructure}}
\tablehead{\colhead{Reference} & \colhead{Associations} & \colhead{Identified CDTGs}}
\startdata
\multirow{24}{*}{\citet{Shank_2022b}} & DTG-27 & 7, 36\\ \cline{2-3}
 & DTG-4 & 26\\ \cline{2-3}
 & DTG-7 & 7\\ \cline{2-3}
 & DTG-12 & 20\\ \cline{2-3}
 & DTG-135 & 11\\ \cline{2-3}
 & DTG-137 & 4\\ \cline{2-3}
 & DTG-156 & 31\\ \cline{2-3}
 & DTG-161 & 24\\ \cline{2-3}
 & DTG-168 & 29\\ \cline{2-3}
 & DTG-170 & 21\\ \cline{2-3}
 & DTG-18 & 2\\ \cline{2-3}
 & DTG-20 & 11\\ \cline{2-3}
 & DTG-22 & 10\\ \cline{2-3}
 & DTG-30 & 8\\ \cline{2-3}
 & DTG-35 & 13\\ \cline{2-3}
 & DTG-38 & 9\\ \cline{2-3}
 & DTG-43 & 38\\ \cline{2-3}
 & DTG-45 & 36\\ \cline{2-3}
 & DTG-50 & 34\\ \cline{2-3}
 & DTG-51 & 1\\ \cline{2-3}
 & DTG-71 & 35\\ \cline{2-3}
 & DTG-79 & 11\\ \cline{2-3}
 & DTG-95 & 8\\ \cline{2-3}
 & DTG-98 & 25\\ \cline{1-3}
\multirow{12}{*}{\citet{Limberg_2021a}} & DTG-11 & 25, 34\\ \cline{2-3}
 & DTG-31 & 2, 21\\ \cline{2-3}
 & DTG-34 & 8, 29\\ \cline{2-3}
 & DTG-4 & 14\\ \cline{2-3}
 & DTG-6 & 3\\ \cline{2-3}
 & DTG-17 & 9\\ \cline{2-3}
 & DTG-19 & 40\\ \cline{2-3}
 & DTG-20 & 33\\ \cline{2-3}
 & DTG-21 & 25\\ \cline{2-3}
 & DTG-28 & 8\\ \cline{2-3}
 & DTG-35 & 29\\ \cline{2-3}
 & DTG-38 & 35\\ \cline{1-3}
\multirow{9}{*}{\citet{Hattori_2022}} & DTC-3 & 1, 4, 16\\ \cline{2-3}
 & DTC-4 & 1, 37\\ \cline{2-3}
 & DTC-16 & 2, 6\\ \cline{2-3}
 & DTC-1 & 14\\ \cline{2-3}
 & DTC-2 & 8\\ \cline{2-3}
 & DTC-15 & 11\\ \cline{2-3}
 & DTC-22 & 17\\ \cline{2-3}
 & DTC-24 & 1\\ \cline{2-3}
 & DTC-29 & 15\\ \cline{1-3}
\multirow{9}{*}{\citet{Shank_2022a}} & DTG-6 & 14, 18\\ \cline{2-3}
 & DTG-4 & 26\\ \cline{2-3}
 & DTG-8 & 31\\ \cline{2-3}
 & DTG-17 & 36\\ \cline{2-3}
 & DTG-27 & 34\\ \cline{2-3}
 & DTG-29 & 11\\ \cline{2-3}
 & DTG-44 & 38\\ \cline{2-3}
 & DTG-50 & 2\\ \cline{2-3}
 & DTG-51 & 21\\ \cline{1-3}
\multirow{8}{*}{\citet{Gudin_2021}} & CDTG-21 & 16, 25\\ \cline{2-3}
 & CDTG-3 & 1\\ \cline{2-3}
 & CDTG-9 & 8\\ \cline{2-3}
 & CDTG-11 & 1\\ \cline{2-3}
 & CDTG-13 & 8\\ \cline{2-3}
 & CDTG-16 & 1\\ \cline{2-3}
 & CDTG-27 & 6\\ \cline{2-3}
 & CDTG-28 & 8\\ \cline{1-3}
\multirow{5}{*}{\citet{Malhan_2022}} & C-3 & 3, 4, 5, 6, 9, 12, 14, 15, 23, 26\\ \cline{2-3}
 & NGC7089 & 12, 16, 31, 32, 39\\ \cline{2-3}
 & Gaia-6 & 3, 6\\ \cline{2-3}
 & Fimbulthul & 2\\ \cline{2-3}
 & Gunnthra & 6\\ 
\citet{Malhan_2022} & Hrid & 34\\ \cline{1-3}
\multirow{5}{*}{\citet{Vasiliev_2021}} & FSR~1758 & 32\\ \cline{2-3}
 & NGC~6235 & 6\\ \cline{2-3}
 & NGC~6356 & 6\\ \cline{2-3}
 & Pfleiderer~2 & 40\\ \cline{2-3}
 & Ryu~879~(RLGC~2) & 20\\ \cline{1-3}
\multirow{4}{*}{\citet{Lovdal_2022}} & 3 & 7, 25, 29\\ \cline{2-3}
 & 1 & 8\\ \cline{2-3}
 & 59 & 17\\ \cline{2-3}
 & 7 & 34\\ \cline{1-3}
\multirow{3}{*}{\citet{Sestito_2019}} & IH & 9, 23, 30, 32\\ \cline{2-3}
 & OH & 15, 30\\ \cline{2-3}
 & P & 29\\ \cline{1-3}
\multirow{3}{*}{\citet{Yuan_2020b}} & DTG-33 & 6\\ \cline{2-3}
 & DTG-40 & 35\\ \cline{2-3}
 & DTG-44 & 26\\ \cline{1-3}
\multirow{2}{*}{\citet{Cordoni_2021}} & Sausage & 8, 17, 29, 31, 34, 35, 36\\ \cline{2-3}
 & Sequoia & 2, 6, 21, 40\\ \cline{1-3}
\multirow{2}{*}{\citet{Monty_2020}} & Sausage & 17, 25, 30, 34\\ \cline{2-3}
 & SeqG1 & 21\\ \cline{1-3}
\multirow{2}{*}{\citet{Roederer_2018}} & A & 1\\ \cline{2-3}
 & C & 1\\ \cline{1-3}
\multirow{1}{*}{\citet{Helmi_2017}} & VelHel-6 & 4\\ \cline{1-3}
\multirow{1}{*}{\citet{Li_2019}} & GL-1 & 9, 12\\ \cline{1-3}
\multirow{1}{*}{\citet{Li_2020}} & GR-2 & 6\\ \cline{1-3}
\multirow{1}{*}{\citet{Myeong_2017}} & Comoving & 11, 26, 30, 34\\ \cline{1-3}
\multirow{1}{*}{\citet{Myeong_2018c}} & Rg5 & 3\\ \cline{1-3}
\enddata
\tablecomments{We draw attention to the associations with CDTGs from \citet{Gudin_2021} and \citet{Roederer_2018} due to their enhancement in $r$-process abundances.}
\end{deluxetable*}

%% file: Tables/cluster_mean_table.tex
\startlongtable
\begin{deluxetable*}{l  r  c  c  c  c  c  c  c}
\tablecaption{CDTG Abundance Means, Dispersions and Inter-Quartile Ranges (IQR) \label{tab:cluster_mean}}
\tablehead{\colhead{Cluster} & \colhead{$N$ Stars} & \colhead{[Fe/H]} & \colhead{[C/Fe]$_{c}$} & \colhead{[Mg/Fe]} & \colhead{[Sr/Fe]} & \colhead{[Y/Fe]} & \colhead{[Ba/Fe]} & \colhead{[Eu/Fe]}}
\startdata
CDTG-$1$ & $17$ & $-2.52\pm0.51$ &$+1.09\pm0.62$ &$+0.34\pm0.28$ &$+0.11\pm0.62$ &$+1.54\pm \dots$ &$+0.41\pm1.04$ &$+0.57\pm0.71$\\
CDTG-$2$ & $12$ & $-2.30\pm0.29$ &$+0.92\pm0.57$ &$+0.37\pm0.16$ &$+0.72\pm0.71$ &$-0.23\pm0.61$ &$+0.12\pm0.91$ &$+0.46\pm0.42$\\
CDTG-$3$ & $11$ & $-2.69\pm0.35$ &$+1.40\pm0.89$ &$+0.42\pm0.06$ &$+0.13\pm0.35$ &$+0.20\pm \dots$ &$+0.62\pm1.48$ &$+0.82\pm0.82$\\
CDTG-$4$ & $10$ & $-2.22\pm0.28$ &$+1.10\pm0.58$ &$+0.31\pm0.15$ &$+0.50\pm0.57$ &$-0.20\pm \dots$ &$+0.57\pm1.16$ &$+0.86\pm0.76$\\
CDTG-$5$ & $10$ & $-2.66\pm0.57$ &$+0.94\pm0.43$ &$+0.50\pm0.26$ &$+0.50\pm0.63$ &$+0.27\pm \dots$ &$-0.22\pm1.00$ &$+0.34\pm0.48$\\
CDTG-$6$ & $10$ & $-2.26\pm0.32$ &$+1.07\pm0.59$ &$+0.41\pm0.13$ &$-0.04\pm0.74$ &$\dots$ &$+0.10\pm1.59$ &$+0.66\pm0.56$\\
CDTG-$7$ & $9$ & $-2.55\pm0.37$ &$+0.80\pm0.16$ &$+0.43\pm0.06$ &$+0.51\pm0.31$ &$\dots$ &$+1.06\pm1.04$ &$+0.48\pm0.84$\\
CDTG-$8$ & $9$ & $-2.27\pm0.52$ &$+0.82\pm0.26$ &$+0.34\pm0.14$ &$+0.36\pm0.62$ &$\dots$ &$-0.04\pm0.80$ &$+0.29\pm0.50$\\
CDTG-$9$ & $9$ & $-2.66\pm0.92$ &$+1.39\pm0.63$ &$+0.44\pm0.30$ &$+0.18\pm1.17$ &$\dots$ &$-0.16\pm1.58$ &$+0.39\pm0.62$\\
CDTG-$10$ & $8$ & $-2.51\pm0.36$ &$+1.38\pm0.45$ &$+0.51\pm0.23$ &$+0.38\pm \dots$ &$\dots$ &$-0.18\pm1.01$ &$+0.22\pm0.40$\\
CDTG-$11$ & $8$ & $-2.20\pm0.35$ &$+0.82\pm0.24$ &$+0.48\pm \dots$ &$+0.41\pm0.40$ &$-0.08\pm0.01$ &$+0.29\pm0.76$ &$+0.94\pm0.36$\\
CDTG-$12$ & $7$ & $-2.71\pm0.62$ &$+1.09\pm0.35$ &$+0.55\pm0.11$ &$+0.21\pm0.65$ &$\dots$ &$-0.13\pm0.99$ &$+0.14\pm0.94$\\
CDTG-$13$ & $7$ & $-1.95\pm0.38$ &$+0.90\pm0.21$ &$+0.69\pm \dots$ &$+1.82\pm0.18$ &$+0.99\pm0.95$ &$+1.43\pm1.00$ &$+1.19\pm0.34$\\
CDTG-$14$ & $7$ & $-2.55\pm0.81$ &$+1.33\pm0.60$ &$+0.38\pm0.29$ &$+0.10\pm0.53$ &$-0.10\pm \dots$ &$+1.36\pm1.38$ &$+1.56\pm \dots$\\
CDTG-$15$ & $6$ & $-3.50\pm0.66$ &$+1.27\pm0.75$ &$+0.47\pm0.07$ &$-0.56\pm0.86$ &$\dots$ &$+0.02\pm1.31$ &$+0.40\pm0.60$\\
CDTG-$16$ & $6$ & $-2.51\pm0.66$ &$+0.89\pm0.43$ &$+0.43\pm0.08$ &$-0.08\pm0.23$ &$\dots$ &$+0.08\pm0.53$ &$+0.86\pm0.72$\\
CDTG-$17$ & $6$ & $-2.21\pm0.31$ &$+1.19\pm0.53$ &$+0.46\pm0.09$ &$+0.18\pm0.45$ &$+0.17\pm \dots$ &$+0.34\pm0.52$ &$+0.39\pm0.19$\\
CDTG-$18$ & $6$ & $-2.81\pm0.71$ &$+1.34\pm0.42$ &$+0.34\pm0.16$ &$+0.33\pm1.09$ &$\dots$ &$-0.49\pm0.77$ &$+0.79\pm0.23$\\
CDTG-$19$ & $5$ & $-2.62\pm0.77$ &$+1.23\pm0.55$ &$+0.54\pm0.21$ &$+0.39\pm0.72$ &$\dots$ &$+0.11\pm1.06$ &$+0.18\pm0.53$\\
CDTG-$20$ & $5$ & $-2.44\pm0.21$ &$+1.12\pm1.04$ &$+0.34\pm0.11$ &$+0.37\pm0.57$ &$\dots$ &$+1.15\pm1.20$ &$+0.48\pm0.12$\\
CDTG-$21$ & $5$ & $-2.34\pm0.08$ &$+1.36\pm0.37$ &$+0.29\pm0.28$ &$-0.01\pm1.40$ &$\dots$ &$+1.03\pm1.07$ &$+0.40\pm0.43$\\
CDTG-$22$ & $5$ & $-2.40\pm0.57$ &$+0.78\pm0.07$ &$+1.17\pm \dots$ &$-0.18\pm \dots$ &$+0.09\pm0.34$ &$-0.21\pm1.02$ &$+0.88\pm \dots$\\
CDTG-$23$ & $5$ & $-2.92\pm0.71$ &$+0.78\pm0.10$ &$+0.90\pm0.50$ &$-0.00\pm0.79$ &$-0.15\pm \dots$ &$-0.34\pm0.98$ &$+0.38\pm \dots$\\
CDTG-$24$ & $5$ & $-1.89\pm1.04$ &$+1.76\pm0.64$ &$+0.80\pm \dots$ &$+1.50\pm0.79$ &$\dots$ &$+1.91\pm0.29$ &$+1.11\pm \dots$\\
CDTG-$25$ & $4$ & $-2.21\pm0.34$ &$+1.65\pm0.70$ &$+0.54\pm \dots$ &$+1.16\pm \dots$ &$-0.32\pm \dots$ &$+0.58\pm \dots$ &$+0.55\pm \dots$\\
CDTG-$26$ & $4$ & $-2.55\pm0.33$ &$+0.84\pm0.28$ &$+0.53\pm \dots$ &$-0.12\pm \dots$ &$-0.20\pm \dots$ &$-0.50\pm0.27$ &$-0.23\pm \dots$\\
CDTG-$27$ & $4$ & $-2.80\pm0.65$ &$+1.17\pm0.33$ &$+0.38\pm0.24$ &$+0.39\pm0.47$ &$+1.73\pm \dots$ &$+0.27\pm0.32$ &$+0.94\pm \dots$\\
CDTG-$28$ & $4$ & $-2.46\pm0.54$ &$+0.90\pm0.21$ &$+0.17\pm \dots$ &$+0.83\pm \dots$ &$+0.24\pm0.17$ &$+0.37\pm0.16$ &$+0.57\pm \dots$\\
CDTG-$29$ & $4$ & $-2.39\pm0.65$ &$+2.75\pm0.81$ &$+0.30\pm \dots$ &$+0.80\pm1.08$ &$\dots$ &$+1.49\pm1.13$ &$+1.39\pm0.71$\\
CDTG-$30$ & $4$ & $-3.44\pm0.52$ &$+2.16\pm1.60$ &$+0.52\pm0.41$ &$+0.67\pm1.41$ &$\dots$ &$+0.69\pm \dots$ &$+1.26\pm \dots$\\
CDTG-$31$ & $4$ & $-2.86\pm0.67$ &$+0.77\pm0.09$ &$+0.31\pm0.25$ &$-0.22\pm0.41$ &$\dots$ &$-0.83\pm0.64$ &$+0.90\pm \dots$\\
CDTG-$32$ & $4$ & $-3.00\pm1.09$ &$+1.88\pm0.41$ &$+0.39\pm0.39$ &$+0.07\pm \dots$ &$\dots$ &$+1.02\pm0.71$ &$+0.25\pm \dots$\\
CDTG-$33$ & $4$ & $-2.63\pm0.44$ &$+0.80\pm0.30$ &$+0.50\pm0.05$ &$+0.12\pm0.32$ &$\dots$ &$-0.41\pm0.21$ &$+0.60\pm0.17$\\
CDTG-$34$ & $3$ & $-2.62\pm0.46$ &$+1.80\pm0.54$ &$+0.44\pm0.17$ &$+0.53\pm \dots$ &$\dots$ &$+1.35\pm0.19$ &$+0.55\pm \dots$\\
CDTG-$35$ & $3$ & $-2.81\pm0.41$ &$+1.32\pm0.60$ &$+0.42\pm \dots$ &$-0.42\pm \dots$ &$+0.15\pm \dots$ &$-0.39\pm \dots$ &$+1.13\pm \dots$\\
CDTG-$36$ & $3$ & $-2.27\pm0.25$ &$+0.75\pm0.03$ &$+0.59\pm0.34$ &$+0.59\pm0.14$ &$\dots$ &$+0.38\pm0.70$ &$+0.72\pm0.13$\\
CDTG-$37$ & $3$ & $-2.60\pm0.04$ &$+1.05\pm0.16$ &$\dots$ &$-0.01\pm \dots$ &$-0.36\pm \dots$ &$-0.01\pm0.01$ &$+0.50\pm \dots$\\
CDTG-$38$ & $3$ & $-2.28\pm0.20$ &$+1.35\pm0.79$ &$+0.36\pm \dots$ &$+0.33\pm0.27$ &$\dots$ &$+0.33\pm0.68$ &$+0.51\pm0.27$\\
CDTG-$39$ & $3$ & $-2.86\pm0.29$ &$+0.91\pm0.22$ &$+0.27\pm0.08$ &$-0.42\pm0.06$ &$\dots$ &$-0.85\pm0.67$ &$-0.05\pm0.47$\\
CDTG-$40$ & $3$ & $-2.20\pm0.19$ &$+1.19\pm0.27$ &$+0.23\pm \dots$ &$+0.59\pm \dots$ &$-0.21\pm \dots$ &$+0.78\pm0.70$ &$+0.86\pm \dots$\\
\cline{0-8}
Biweight (CDTG mean): & & $-2.52\pm0.05$ & $+1.12\pm0.05$ & $+0.42\pm0.02$ & $+0.27\pm0.07$ & $-0.03\pm0.08$ & $+0.29\pm0.11$ & $+0.60\pm0.06$\\
Biweight (CDTG std): & & $+0.47\pm0.04$ & $+0.43\pm0.04$ & $\dots$ & $\dots$ & $\dots$ & $\dots$ & $\dots$\\
IQR (CDTG mean): & & $\boldsymbol{0.42}$ & $\boldsymbol{0.46}$ & $\boldsymbol{0.18}$ & $\boldsymbol{0.52}$ & $\boldsymbol{0.43}$ & $\boldsymbol{0.88}$ & $\boldsymbol{0.48}$\\
IQR (CDTG std): & & $0.34$ & $0.35$ & $NaN$ & $NaN$ & $NaN$ & $NaN$ & $NaN$\\
\cline{0-8}
Biweight (Final): & & $-2.57\pm0.59$ & $+1.13\pm0.63$ & $+0.43\pm0.21$ & $+0.17\pm0.71$ & $-0.06\pm0.47$ & $+0.17\pm1.13$ & $+0.51\pm0.65$\\
IQR (Final): & & $\boldsymbol{0.80}$ & $\boldsymbol{0.97}$ & $\boldsymbol{0.26}$ & $\boldsymbol{0.86}$ & $\boldsymbol{0.47}$ & $\boldsymbol{1.62}$ & $\boldsymbol{0.80}$\\
\enddata
\tablecomments{The first section of the table lists the biweight location and scale of the abundances for each of the CDTGs. The second section of the table lists the mean and the standard error of the mean (using biweight estimates) for both the location and scale of the abundances of the CDTGs, along with the IQR of the abundances of the CDTGs. The third section of the table lists the biweight location and scale of the Final Sample for each of the abundances, along with the IQR for each of the abundances in the Final Sample.}\end{deluxetable*}

%% file: Tables/binomial_probability_table_full.tex
\begin{deluxetable*}{cccccc}[t!]
\tabletypesize{\footnotesize}
\tablecaption{CDTG Elemental-Abundance Statistics\label{tab:binomial_probability_table_full}}
\tablehead{\colhead{Abundance} & \colhead{$\#$ CDTGs} & \colhead{$  N < 0.50$, $0.33$, $0.25$} & \colhead{IEAD Probabilities} & \colhead{GEAD Probabilities} & \colhead{OEAD Probability}}
\startdata
 &  &  & All CDTGs &  &  \\
\hline
\text{[Fe/H]} & $40$ &  $20$, $16$, $15$ & $56.3\%$, $21.7\%$, $\phantom{0}5.4\%$ & $\phantom{0}5.1\%$ & \hfil\multirow{4}{*}{$\phantom{0}0.3\%$}\hfill \\
\text{[C/Fe]$_\text{c}$} & $40$ & $22$, $16$, $12$ & $31.8\%$, $21.7\%$, $28.5\%$ & $12.9\%$ & \\
\text{[Mg/Fe]} & $28$ & $13$, $10$, $\phantom{1}6$ & $71.4\%$, $44.9\%$, $73.6\%$ & $41.0\%$ & \\
\multicolumn{2}{c}{FEAD Probabilities} & & $28.0\%$, $\phantom{0}4.3\%$, $\phantom{0}4.9\%$ & & \\
\hline
 &  &  & Group I CDTGs &  &  \\
\hline
\text{[Fe/H]} & $16$ &  $\phantom{0}8$, $\phantom{0}5$, $\phantom{0}5$ & $59.8\%$, $65.0\%$, $37.0\%$ & $31.5\%$ & \hfil\multirow{4}{*}{$12.0\%$}\hfill \\
\text{[C/Fe]$_\text{c}$} & $16$ & $\phantom{0}7$, $\phantom{0}2$, $\phantom{0}1$ & $77.3\%$, $98.5\%$, $99.0\%$ & $76.9\%$ & \\
\text{[Mg/Fe]} & $\phantom{0}7$ & $\phantom{0}4$, $\phantom{0}1$, $\phantom{0}0$ & $50.0\%$, $93.9\%$, $100.0\%$ & $49.6\%$ & \\
\multicolumn{2}{c}{FEAD Probabilities} & & $39.5\%$, $91.5\%$, $75.0\%$ & & \\
\hline
 &  &  & Group II CDTGs &  &  \\
\hline
\text{[Fe/H]} & $\phantom{0}9$ &  $\phantom{0}5$, $\phantom{0}2$, $\phantom{0}1$ & $47.8\%$, $85.2\%$, $92.5\%$ & $47.8\%$ & \hfil\multirow{4}{*}{$< 0.1\%$}\hfill \\
\text{[C/Fe]$_\text{c}$} & $\phantom{0}9$ & $\phantom{0}7$, $\phantom{0}7$, $\phantom{0}5$ & $\phantom{1}9.0\%$, $\phantom{1}0.8\%$, $\phantom{1}4.9\%$ & $\phantom{0}0.6\%$ & \\
\text{[Mg/Fe]} & $\phantom{0}5$ & $\phantom{0}3$, $\phantom{0}3$, $\phantom{0}3$ & $50.0\%$, $20.5\%$, $10.4\%$ & $10.4\%$ & \\
\multicolumn{2}{c}{FEAD Probabilities} & & $\phantom{0}5.6\%$, $\phantom{0}0.7\%$, ~$\phantom{0}2.0\%$ & & \\
\enddata
\tablecomments{The Individual Elemental-Abundance Dispersion (IEAD) probabilities represent the 
binomial probabilities for each element for the levels $v$ =  0.50, 0.33, and 0.25, respectively.
The Full Elemental-Abundance Dispersion (FEAD) probabilities represent the probabilities
(across {\it all} elements) for the levels $v$ =  0.50, 0.33, and 0.25, respectively.
The Global Elemental-Abundance Dispersion (GEAD) probabilities represent the probabilities 
for the triplet of CDF levels for each element.
The Overall Elemental-Abundance Dispersion (OEAD) probability represents the probability
(across {\it all} elements) resulting from random draws from the full CDF.}
\end{deluxetable*}

%% file: Tables/initial_data_description_table.tex
\startlongtable
\begin{deluxetable*}{c  l  l  l}
\tablecaption{Description of the Initial Sample of CEMP stars \label{tab:initial_data_descript}}
\tablehead{\colhead{Column} & \colhead{Field} & \colhead{Unit} & \colhead{Description}}
\startdata
1 & Name & $-$ & The name of the star as given by the Reference\\
2 & Source ID & $-$ & The Gaia EDR3 Source ID of the star\\
3 & RA & (J2000) & The Right Ascension of the star given in hours:minutes:seconds\\
4 & DEC & (J2000) & The Declination of the star given in degrees:minutes:seconds\\
5 & $V_{\rm mag}$ & $-$ & The $V$ magnitude of the star as given by the $V_{\rm mag}$ Reference\\
6 & $G_{\rm mag}$ & $-$ & The Gaia $G$ mean magnitude of the star as given by the Gaia Source ID\\
7 & $G_{\rm BP} - G_{\rm RP}$ & $-$ & The Gaia BP $-$ RP color mean magnitude of the star as given by the Gaia Source ID\\
8 & $V_{\rm mag}$ (Gaia) & $-$ & The $V$ magnitude of the star as determined by the transformations from G mag to V mag\\
 &  &  & using $V$ $=$ $G$ $+$ $0.02704$ $-$ $0.01424*(\rm{BP}-\rm{RP})$ $+$ $0.2156*(\rm{BP}-\rm{RP})^2$ $-$\\
 &  &  & $0.01426(\rm{BP}-\rm{RP})^3$ given by \citet{Riello_2021}\\
9 & RV & (km s$^{-1}$) & The radial velocity as given by RV Reference\\
10 & Error & (km s$^{-1}$) & The radial velocity error as given by RV Reference\\
11 & RV$_{\rm Gaia}$ & (km s$^{-1}$) & The radial velocity as given by the Gaia Source ID\\
12 & Error & (km s$^{-1}$) & The radial velocity error as given by the Gaia Source ID\\
13 & Parallax & (mas) & The parallax as given by the Gaia Source ID\\
14 & Error & (mas) & The parallax error as given by the Gaia Source ID\\
15 & Distance & (kpc) & The inverse parallax distance (1/Parallax)\\
16 & Error & (kpc) & The inverse parallax distance error (Parallax$_{error}$/(Parallax$^2$))\\
19 & Relative Error & $-$ & The relative error of the corrected distance as given by Gaia\\
20 & Distance BJ21 & (kpc) & The 50 percentile distance as given by \citet{Bailer_Jones_2021} based on the Gaia Source ID\\
21 & Error & (kpc) & The 50 percentile error as estimated by the 84 percentile distance and the 16 percentile\\
 &  &  & distance as given by \citet{Bailer_Jones_2021} based on the Gaia Source ID ((dist84-dist16)/2)\\
22 & Relative Error & $-$ & The relative error of the 50 percentile distance as given by \citet{Bailer_Jones_2021} based on\\
 &  &  & the Gaia Source ID\\
23 & Distance StarHorse & (kpc) & The 50 percentile distance as given by \citet{Anders_2021} based on the Gaia Source ID\\
24 & Error & (kpc) & The 50 percentile error as estimated by the 84 percentile distance and the 16 percentile\\
 &  &  & distance as given by \citet{Anders_2021} based on the Gaia Source ID ((dist84-dist16)/2)\\
25 & Relative Error & $-$ & The relative error of the 50 percentile distance as given by \citet{Anders_2021} based on the\\
 &  &  & Gaia Source ID\\
26 & PM$_{\rm RA}$ & (mas yr$^{-1}$) & The proper motion in the Right Ascension as given by the Gaia Source ID\\
27 & Error & (mas yr$^{-1}$) & The proper motion error in the Right Ascension as given by the Gaia Source ID\\
28 & PM$_{\rm DEC}$ & (mas yr$^{-1}$) & The proper motion in the Declination as given by the Gaia Source ID\\
29 & Error & (mas yr$^{-1}$) & The proper motion error in the Declination as given by the Gaia Source ID\\
30 & Correlation Coefficient & $-$ & The correlation coefficient between the proper motion in Right Ascension and the proper\\
 &  &  & motion in Declination as given by the Gaia Source ID\\
31 & T$_{\rm eff}$ & (K) & The effective temperature of the star as given by the Reference\\
32 & log \textit{g} & (cgs) & The surface gravity of the star as given by the Reference\\
33 & [Fe/H] & $-$ & The metallicity of the star as given by (log $\epsilon$(Fe) - log $\epsilon$(Fe)$_\odot$)\\
 &  &  & (Solar value of 7.5 taken from \citet{Asplund_2009})\\
34 & log $\epsilon$(Fe) & $-$ & The logarithmic iron abundance of the star as given by the Reference\\
35 & [C/Fe] & $-$ & The carbon abundance ratio of the star as given by (log $\epsilon$(C) - log\\
 &  &  & $\epsilon$(C)$_\odot$ - [Fe/H]) (Solar value of 8.43 taken from \citet{Asplund_2009})\\
36 & log $\epsilon$(C) & $-$ & The logarithmic carbon abundance of the star as given by the Reference\\
37 & [C/Fe]$_{c}$ & $-$ & The carbon abundance ratio corrected for evolutionary effects from \citet{Placco_2014}\\
38 & AC$_{c}$ & $-$ & The absolute carbon corrected for evolutionary effects from \citet{Placco_2014} ([C/Fe]$_{c}$\\
 &  &  & + [Fe/H] + log $\epsilon$(C)$_\odot$) (Solar value of 8.43 taken from \citet{Asplund_2009})\\
39 & CARDET & $-$ & Flag with ``D" if the carbon abundance ratio ([C/Fe]) is detected by Reference and ``U"\\
 &  &  & if an upper limit by Reference and ``L" if a lower limit by Reference and ``X" if none is\\
 &  &  & detected by Reference\\
40 & [Mg/Fe] & $-$ & The magnesium abundance ratio of the star as given by (log $\epsilon$(Mg) - log\\
 &  &  & $\epsilon$(Mg)$_\odot$ - [Fe/H]) (Solar value of 7.60 taken from \citet{Asplund_2009})\\
41 & log $\epsilon$(Mg) & $-$ & The logarithmic magnesium abundance of the star as given by the Reference\\
42 & MAGDET & $-$ & Flag with ``D" if the magnesium abundance ratio ([Mg/Fe]) is detected by Reference and ``U"\\
 &  &  & if an upper limit by Reference and ``L" if a lower limit by Reference and ``X" if none is\\
 &  &  & detected by Reference\\
43 & [Sr/Fe] & $-$ & The strontium abundance ratio of the star as given by (log $\epsilon$(Sr) - log\\
 &  &  & $\epsilon$(Sr)$_\odot$ - [Fe/H]) (Solar value of 2.87 taken from \citet{Asplund_2009})\\
44 & log $\epsilon$(Sr) & $-$ & The logarithmic strontium abundance of the star as given by the Reference\\
45 & STRDET & $-$ & Flag with ``D" if the strontium abundance ratio ([Sr/Fe]) is detected by Reference and ``U"\\
 &  &  & if an upper limit by Reference and ``L" if a lower limit by Reference and ``X" if none is\\
 &  &  & detected by Reference\\
46 & [Y/Fe] & $-$ & The yttrium abundance ratio of the star as given by (log $\epsilon$(Y) - log\\
 &  &  & $\epsilon$(Y)$_\odot$ - [Fe/H]) (Solar value of 2.21 taken from \citet{Asplund_2009})\\
47 & log $\epsilon$(Y) & $-$ & The logarithmic yttrium abundance of the star as given by the Reference\\
48 & YTTDET & $-$ & Flag with ``D" if the yttrium abundance ratio ([Y/Fe]) is detected by Reference and ``U"\\
 &  &  & if an upper limit by Reference and ``L" if a lower limit by Reference and ``X" if none is\\
 &  &  & detected by Reference\\
49 & [Ba/Fe] & $-$ & The barium abundance ratio of the star as given by (log $\epsilon$(Ba) - log\\
 &  &  & $\epsilon$(Ba)$_\odot$ - [Fe/H]) (Solar value of 2.18 taken from \citet{Asplund_2009})\\
50 & log $\epsilon$(Ba) & $-$ & The logarithmic barium abundance of the star as given by the Reference\\
51 & BARDET & $-$ & Flag with ``D" if the barium abundance ratio ([Ba/Fe]) is detected by Reference and ``U"\\
 &  &  & if an upper limit by Reference and ``L" if a lower limit by Reference and ``X" if none is\\
 &  &  & detected by Reference\\
52 & [Eu/Fe] & $-$ & The europium abundance ratio of the star as given by (log $\epsilon$(Eu) - log\\
 &  &  & $\epsilon$(Eu)$_\odot$ - [Fe/H]) (Solar value of 0.52 taken from \citet{Asplund_2009})\\
53 & log $\epsilon$(Eu) & $-$ & The logarithmic europium abundance of the star as given by the Reference\\
54 & EURDET & $-$ & Flag with ``D" if the europium abundance ratio ([Eu/Fe]) is detected by Reference and ``U"\\
 &  &  & if an upper limit by Reference and ``L" if a lower limit by Reference and ``X" if none is\\
 &  &  & detected by Reference\\
55 & Class & $-$ & The class for the star, as given by Table 1\\
56 & Reference & $-$ & The Reference for the star\\
57 & $V_{\rm mag}$ Reference & $-$ & The Reference for the \textit{V} magnitude of the star\\
58 & Distance AGAMA & $-$ & The Reference for the distance used in AGAMA (StarHorse prioritized over BJ21 unless\\
 &  &  & StarHorse distance has relative error greater than 0.3, if both have a relative error greater\\
 &  &  & than 0.3 we adopt no distance estimate)\\
\enddata
\end{deluxetable*}

%% file: Tables/final_data_description_table.tex
\startlongtable
\begin{deluxetable*}{c  l  l  l}
\tablecaption{Description of the Final Sample \label{tab:final_data_descript}}
\tablehead{\colhead{Column} & \colhead{Field} & \colhead{Unit} & \colhead{Description}}
\startdata
1 & Name & $-$ & The name of the star as given by the Reference\\
2 & Source ID & $-$ & The Gaia EDR3 Source ID of the star\\
3 & RA & (J2000) & The Right Ascension of the star given in hours:minutes:seconds\\
4 & DEC & (J2000) & The Declination of the star given in degrees:minutes:seconds\\
5 & $V_{\rm mag}$ & $-$ & The $V$ magnitude of the star as given by the $V_{\rm mag}$ Reference\\
6 & $G_{\rm mag}$ & $-$ & The Gaia $G$ mean magnitude of the star as given by the Gaia Source ID\\
7 & $G_{\rm BP} - G_{\rm RP}$ & $-$ & The Gaia BP $-$ RP color mean magnitude of the star as given by the Gaia Source ID\\
8 & $V_{\rm mag}$ (Gaia) & $-$ & The $V$ magnitude of the star as determined by the transformations from G mag to\\
 &  &  & V mag using $V$ $=$ $G$ $+$ $0.02704$ $-$ $0.01424*(\rm{BP}-\rm{RP})$ $+$ $0.2156*(\rm{BP}-\rm{RP})^2$ $-$\\
 &  &  & $0.01426(\rm{BP}-\rm{RP})^3$ given by \citet{Riello_2021}\\
9 & RV & (km s$^{-1}$) & The radial velocity as given by RV Reference\\
10 & Error & (km s$^{-1}$) & The radial velocity error as given by RV Reference\\
11 & RV$_{\rm Gaia}$ & (km s$^{-1}$) & The radial velocity as given by the Gaia Source ID\\
12 & Error & (km s$^{-1}$) & The radial velocity error as given by the Gaia Source ID\\
13 & Parallax & (mas) & The parallax as given by the Gaia Source ID\\
14 & Error & (mas) & The parallax error as given by the Gaia Source ID\\
15 & Distance & (kpc) & The inverse parallax distance (1/Parallax)\\
16 & Error & (kpc) & The inverse parallax distance error (Parallax$_{error}$/(Parallax$^2$))\\
17 & Relative Error & $-$ & The relative error of the corrected distance as given by Gaia\\
18 & Distance BJ21 & (kpc) & The 50 percentile distance as given by \citet{Bailer_Jones_2021} based on\\
 & & & the Gaia Source ID\\
19 & Error & (kpc) & The 50 percentile error as estimated by the 84 percentile distance and the 16\\
 &  &  & percentile distance as given by \citet{Bailer_Jones_2021} based on the Gaia Source ID\\
 &  &  & ((dist84-dist16)/2)\\
20 & Relative Error & $-$ & The relative error of the 50 percentile distance as given by \citet{Bailer_Jones_2021}\\
 &  &  & based on the Gaia Source ID\\
21 & Distance StarHorse & (kpc) & The 50 percentile distance as given by \citet{Anders_2021} based on the Gaia Source ID\\
22 & Error & (kpc) & The 50 percentile error as estimated by the 84 percentile distance and the 16 percentile\\
 &  &  & distance as given by \citet{Anders_2021} based on the Gaia Source ID ((dist84-dist16)/2)\\
23 & Relative Error & $-$ & The relative error of the 50 percentile distance as given by \citet{Anders_2021} based on\\
 &  &  & the Gaia Source ID\\
24 & PM$_{\rm RA}$ & (mas yr$^{-1}$) & The proper motion in the Right Ascension as given by the Gaia Source ID\\
25 & Error & (mas yr$^{-1}$) & The proper motion error in the Right Ascension as given by the Gaia Source ID\\
26 & PM$_{\rm DEC}$ & (mas yr$^{-1}$) & The proper motion in the Declination as given by the Gaia Source ID\\
27 & Error & (mas yr$^{-1}$) & The proper motion error in the Declination as given by the Gaia Source ID\\
28 & Correlation Coefficient & $-$ & The correlation coefficient between the\\
 &  &  & proper motion in Right Ascension and the proper motion\\
 &  &  & in Declination as given by the Gaia Source ID\\
29 & T$_{\rm eff}$ & (K) & The effective temperature of the star as given by the Reference\\
30 & log \textit{g} & (cgs) & The surface gravity of the star as given by the Reference\\
31 & [Fe/H] & $-$ & The metallicity of the star as given by (log $\epsilon$(Fe) - log $\epsilon$(Fe)$_\odot$)\\
 &  &  & (Solar value of 7.5 taken from \citet{Asplund_2009})\\
32 & log $\epsilon$(Fe) & $-$ & The logarithmic iron abundance of the star as given by the Reference\\
33 & [C/Fe] & $-$ & The carbon abundance ratio of the star as given by (log $\epsilon$(C) - log\\
 &  &  & $\epsilon$(C)$_\odot$ - [Fe/H]) (Solar value of 8.43 taken from \citet{Asplund_2009})\\
34 & log $\epsilon$(C) & $-$ & The logarithmic carbon abundance of the star as given by the Reference\\
35 & [C/Fe]$_{c}$ & $-$ & The carbon abundance ratio corrected for evolutionary effects from \citet{Placco_2014}\\
36 & AC$_{c}$ & $-$ & The absolute carbon corrected for evolutionary effects from \citet{Placco_2014} ([C/Fe]$_{c}$\\
 &  &  & + [Fe/H] + log $\epsilon$(C)$_\odot$) (Solar value of 8.43 taken from \citet{Asplund_2009})\\
37 & CARDET & $-$ & Flag with ``D" if the carbon abundance ratio ([C/Fe]) is detected by Reference and ``U"\\
 &  &  & if an upper limit by Reference and ``L" if a lower limit by Reference and ``X" if none is\\
 &  &  & detected by Reference\\
38 & [Mg/Fe] & $-$ & The magnesium abundance ratio of the star as given by (log $\epsilon$(Mg) - log\\
 &  &  & $\epsilon$(Mg)$_\odot$ - [Fe/H]) (Solar value of 7.60 taken from \citet{Asplund_2009})\\
39 & log $\epsilon$(Mg) & $-$ & The logarithmic magnesium abundance of the star as given by the Reference\\
40 & MAGDET & $-$ & Flag with ``D" if the magnesium abundance ratio ([Mg/Fe]) is detected by Reference and ``U"\\
 &  &  & if an upper limit by Reference and ``L" if a lower limit by Reference and ``X" if none is\\
 &  &  & detected by Reference\\
41 & [Sr/Fe] & $-$ & The strontium abundance ratio of the star as given by (log $\epsilon$(Sr) - log\\
 &  &  & $\epsilon$(Sr)$_\odot$ - [Fe/H]) (Solar value of 2.87 taken from \citet{Asplund_2009})\\
42 & log $\epsilon$(Sr) & $-$ & The logarithmic strontium abundance of the star as given by the Reference\\
43 & STRDET & $-$ & Flag with ``D" if the strontium abundance ratio ([Sr/Fe]) is detected by Reference and ``U"\\
 &  &  & if an upper limit by Reference and ``L" if a lower limit by Reference and ``X" if none is\\
 &  &  & detected by Reference\\
44 & [Y/Fe] & $-$ & The yttrium abundance ratio of the star as given by (log $\epsilon$(Y) - log\\
 &  &  & $\epsilon$(Y)$_\odot$ - [Fe/H]) (Solar value of 2.21 taken from \citet{Asplund_2009})\\
45 & log $\epsilon$(Y) & $-$ & The logarithmic yttrium abundance of the star as given by the Reference\\
46 & YTTDET & $-$ & Flag with ``D" if the yttrium abundance ratio ([Y/Fe]) is detected by Reference and ``U"\\
 &  &  & if an upper limit by Reference and ``L" if a lower limit by Reference and ``X" if none is\\
 &  &  & detectedby Reference\\
47 & [Ba/Fe] & $-$ & The barium abundance ratio of the star as given by (log $\epsilon$(Ba) - log\\
 &  &  & $\epsilon$(Ba)$_\odot$ - [Fe/H]) (Solar value of 2.18 taken from \citet{Asplund_2009})\\
48 & log $\epsilon$(Ba) & $-$ & The logarithmic barium abundance of the star as given by the Reference\\
49 & BARDET & $-$ & Flag with ``D" if the barium abundance ratio ([Ba/Fe]) is detected by Reference and ``U"\\
 &  &  & if an upper limit by Reference and ``L" if a lower limit by Reference and ``X" if none is\\
 &  &  & detected by Reference\\
50 & [Eu/Fe] & $-$ & The europium abundance ratio of the star as given by (log $\epsilon$(Eu) - log\\
 &  &  & $\epsilon$(Eu)$_\odot$ - [Fe/H]) (Solar value of 0.52 taken from \citet{Asplund_2009})\\
51 & log $\epsilon$(Eu) & $-$ & The logarithmic europium abundance of the star as given by the Reference\\
52 & EURDET & $-$ & Flag with ``D" if the europium abundance ratio ([Eu/Fe]) is detected by Reference and ``U"\\
 &  &  & if an upper limit by Reference and ``L" if a lower limit by Reference and ``X" if none is\\
 &  &  & detected by Reference\\
53 & Class & $-$ & The class for the star, as given by Table 1\\
54 & Reference & $-$ & The Reference for the star\\
55 & $V_{\rm mag}$ Reference & $-$ & The Reference for the \textit{V} magnitude of the star\\
56 & Distance AGAMA & $-$ & The Reference for the distance used in AGAMA (StarHorse prioritized over BJ21 unless\\
 &  &  & StarHorse distance has relative error greater than 0.3, if both have a relative error greater\\
 &  &  & than 0.3 we adopt no distance estimate)\\
57 & (v$_{\rm r}$,v$_{\phi}$,v$_{\rm z}$)  & (km s$^{-1}$) & The cylindrical velocities of the star as given by AGAMA\\
58 & Error & (km s$^{-1}$) & The cylindrical velocity errors of the star as given by Monte Carlo sampling through AGAMA\\
59 & (J$_{\rm r}$,J$_{\phi}$,J$_{\rm z}$) & (kpc km s$^{-1}$) & The cylindrical actions of the star as given by AGAMA\\
60 & Error & (kpc km s$^{-1}$) & The cylindrical action errors of the star as given by Monte Carlo sampling through AGAMA\\
61 & Energy & (km$^{2}$ s$^{-2}$) & The orbital energy of the star as given by AGAMA\\
62 & Error  & (km$^{2}$ s$^{-2}$) & The orbital energy error of the star as given by Monte Carlo sampling through AGAMA\\
63 & $r_{\rm peri}$ & (kpc) & The Galactic pericentric distance of the star as given by AGAMA\\
64 & Error & (kpc) & The Galactic pericentric distance error of the star as given by Monte Carlo sampling through\\
 &  &  & AGAMA\\
65 & $r_{\rm apo}$ & (kpc) & The Galactic apocentric distance of the star as given by AGAMA\\
66 & Error & (kpc) & The Galactic apocentric distance error of the star as given by Monte Carlo\\
 &  &  & sampling through AGAMA\\
67 & Z$_{\rm max}$ & (kpc) & The maximum height above the Galactic plane of the star as given by AGAMA\\
68 & Error & (kpc) & The maximum height above the Galactic plane error of the star as given by Monte Carlo\\
 &  &  & sampling through AGAMA\\
69 & Eccentricity & $-$ & The eccentricity of the star given by ($r_{\rm{apo}}$ $-$ $r_{\rm{peri}}$)/($r_{\rm{apo}}$ $+$\\
 &  &  & $r_{\rm{peri}}$) through AGAMA\\
70 & Error & $-$ & The eccentricity error of the star as given by Monte Carlo sampling through AGAMA\\
\enddata
\end{deluxetable*}

%% file: Tables/group1_cluster_stellar_results_stub_table.tex
\begin{deluxetable*}{l  c  c  c  c  c  c  c  c}
\tablecaption{Group I CDTGs Identified by \texttt{HDBSCAN} \label{tab:g1_cluster_results_stub}}
\tablehead{\colhead{NAME} & \colhead{[Fe/H]} & \colhead{[C/Fe]$_{c}$} & \colhead{[Mg/Fe]} & \colhead{[Sr/Fe]} & \colhead{[Y/Fe]} & \colhead{[Ba/Fe]} & \colhead{[Eu/Fe]}}
\startdata
\multicolumn{8}{c}{CDTG$-1-$GI} \\
\multicolumn{8}{c}{Structure: Unassigned Structure} \\
\multicolumn{8}{c}{Group Assoc: Sequoia: \citet{Cordoni_2021}} \\
\multicolumn{8}{c}{Group Assoc: Fimbulthul: \citet{Malhan_2022}} \\
\multicolumn{8}{c}{Group Assoc: DTG-50: \citet{Shank_2022a}} \\
\multicolumn{8}{c}{Group Assoc: DTG-18: \citet{Shank_2022b}} \\
\multicolumn{8}{c}{Group Assoc: DTG-47: \citet{Shank_2022b}} \\
\multicolumn{8}{c}{Stellar Assoc: BS16543-97 (DTG-31: \citealt{Limberg_2021a})} \\
\multicolumn{8}{c}{Stellar Assoc: BPS BS 16543-0097 (DTC-16: \citealt{Hattori_2022})} \\
\multicolumn{8}{c}{Stellar Assoc: CS22964-161 (DTG-31: \citealt{Limberg_2021a})} \\
\multicolumn{8}{c}{Stellar Assoc: CS22964-161 (DTG-31: \citealt{Limberg_2021a}) } \\
\multicolumn{8}{c}{Globular Assoc: No Globular Associations} \\
\multicolumn{8}{c}{Dwarf Galaxy Assoc: No Dwarf Galaxy Associations} \\
BPS CS 29497$-$0030 & $-2.52$ & $+2.37$ & $+0.35$ & $+1.30$ & $\dots$ & $+2.75$ & $+1.71$ \\
BPS BS 16543$-$0097 & $-2.48$ & $+1.38$ & $+0.66$ & $+1.08$ & $\dots$ & $+0.43$ & $+0.97$ \\
HE 1343$-$0640 & $-1.90$ & $+0.73$ & $+0.35$ & $+0.78$ & $\dots$ & $+0.66$ & $\dots$ \\
HE 1434$-$1442 & $-2.39$ & $+2.09$ & $+0.29$ & $\dots$ & $\dots$ & $+1.19$ & $\dots$ \\
Pristine J229.1219$+$00.9089 & $-2.25$ & $+1.74$ & $+0.16$ & $+0.92$ & $\dots$ & $\dots$ & $\dots$ \\
GALAH 170530003601365 & $-1.14$ & $+0.92$ & $\dots$ & $+1.30$ & $+1.02$ & $+1.12$ & $+0.29$ \\
BD$+$19 3109 & $-2.23$ & $+2.51$ & $\dots$ & $\dots$ & $\dots$ & $+1.73$ & $+0.72$ \\
BPS CS 22964$-$0161A & $-2.37$ & $+1.58$ & $\dots$ & $\dots$ & $\dots$ & $+1.36$ & $\dots$ \\
BPS CS 22964$-$0161B & $-2.39$ & $+1.40$ & $\dots$ & $\dots$ & $\dots$ & $+1.30$ & $\dots$ \\
BPS CS 22948$-$0027 & $-2.47$ & $+2.58$ & $+0.29$ & $+0.95$ & $\dots$ & $+2.21$ & $+1.87$ \\
BPS CS 22881$-$0036 & $-2.37$ & $+1.99$ & $+0.57$ & $+1.06$ & $\dots$ & $+2.03$ & $+0.58$ \\
$\boldsymbol{\mu \pm \sigma ([X/Y])}$ & $\boldsymbol{-2.37\pm0.15}$ & $\boldsymbol{+1.77\pm0.64}$ & $\boldsymbol{+0.33\pm0.18}$ & $\boldsymbol{+1.05\pm0.19}$ & $\boldsymbol{+1.02\pm \dots}$ & $\boldsymbol{+1.44\pm0.71}$ & $\boldsymbol{+0.96\pm0.64}$\\
\enddata
\tablecomments{$\boldsymbol{\mu}$ and $\boldsymbol{\sigma}$ represent the biweight estimates of the location and scale for the abundances in the CDTG.}\tablecomments{This table is a stub; the full table is available in the electronic edition.}\end{deluxetable*}

%% file: Tables/g1_cluster_orbital_table.tex
\clearpage
\startlongtable
\begin{deluxetable*}{l  r  c  c  r  r}
\tablecaption{Group I Cluster Dynamical Parameters Determined by \texttt{AGAMA} \label{tab:g1_cluster_orbit}}
\tablehead{\colhead{Cluster} & \colhead{$N$ Stars} & \colhead{($\langle$v$_{\text{r}}\rangle$,$\langle$v$_{\phi}\rangle$,$\langle$v$_{\text{z}}\rangle$)} & \colhead{($\langle$J$_{\text{r}}\rangle$,$\langle$J$_{\phi}\rangle$,$\langle$J$_{\text{z}}\rangle$)} & \colhead{$\langle$E$\rangle$} & \colhead{$\langle$ecc$\rangle$}\\
\colhead{} & \colhead{} & \colhead{($\sigma_{\langle\text{v}_{\text{r}}\rangle}$,$\sigma_{\langle\text{v}_{\phi}\rangle}$,$\sigma_{\langle\text{v}_{\text{z}}\rangle}$)} & \colhead{($\sigma_{\langle\text{J}_{\text{r}}\rangle}$,$\sigma_{\langle\text{J}_{\phi}\rangle}$,$\sigma_{\langle\text{J}_{\text{z}}\rangle}$)} & \colhead{$\sigma_{\langle\text{E}\rangle}$} & \colhead{$\sigma_{\langle\text{ecc}\rangle}$}\\
\colhead{} & \colhead{} & \colhead{(km s$^{-1}$)} & \colhead{(kpc km s$^{-1}$)} & \colhead{(10$^{5}$ km$^{2}$ s$^{-2}$)} & \colhead{}}
\startdata
CDTG$-1-$GI & $11$ & ($-22.4$,$-93.1$,$8.4$) & ($286.4$,$-634.0$,$84.8$) & $-1.805$ & $0.61$ \\
 & & ($86.1$,$32.4$,$56.4$) & ($132.1$,$197.4$,$26.6$) & $0.035$ & $0.13$ \\
CDTG$-2-$GI & $8$ & ($9.9$,$155.4$,$12.4$) & ($184.6$,$1280.6$,$69.5$) & $-1.648$ & $0.42$ \\
 & & ($84.4$,$19.5$,$46.6$) & ($61.3$,$62.0$,$38.6$) & $0.029$ & $0.07$ \\
CDTG$-3-$GI & $7$ & ($-17.9$,$82.1$,$-19.7$) & ($392.9$,$686.7$,$116.0$) & $-1.723$ & $0.67$ \\
 & & ($87.8$,$17.6$,$43.1$) & ($38.9$,$73.8$,$40.0$) & $0.026$ & $0.03$ \\
CDTG$-4-$GI & $6$ & ($12.3$,$73.9$,$-133.1$) & ($197.1$,$648.9$,$563.0$) & $-1.630$ & $0.47$ \\
 & & ($34.3$,$1.6$,$13.3$) & ($50.1$,$92.5$,$116.6$) & $0.060$ & $0.04$ \\
CDTG$-5-$GI & $6$ & ($-10.2$,$188.8$,$-8.9$) & ($31.8$,$1543.6$,$40.6$) & $-1.641$ & $0.17$ \\
 & & ($16.1$,$2.3$,$30.9$) & ($6.0$,$48.5$,$30.2$) & $0.027$ & $0.02$ \\
CDTG$-6-$GI & $6$ & ($61.5$,$40.2$,$-25.7$) & ($417.0$,$221.0$,$1338.0$) & $-1.441$ & $0.50$ \\
 & & ($163.0$,$65.8$,$165.9$) & ($259.5$,$191.1$,$140.4$) & $0.063$ & $0.13$ \\
CDTG$-7-$GI & $5$ & ($-56.6$,$-137.3$,$-129.6$) & ($168.9$,$-918.9$,$826.6$) & $-1.482$ & $0.38$ \\
 & & ($87.5$,$48.7$,$67.9$) & ($131.5$,$170.5$,$43.4$) & $0.006$ & $0.14$ \\
CDTG$-8-$GI & $5$ & ($-16.0$,$45.8$,$30.6$) & ($448.1$,$373.5$,$96.8$) & $-1.799$ & $0.78$ \\
 & & ($77.7$,$4.8$,$43.2$) & ($23.3$,$85.5$,$20.6$) & $0.011$ & $0.04$ \\
CDTG$-9-$GI & $5$ & ($-66.1$,$18.4$,$56.8$) & ($211.7$,$130.2$,$2001.7$) & $-1.308$ & $0.32$ \\
 & & ($127.8$,$38.2$,$291.2$) & ($257.9$,$327.0$,$30.3$) & $0.019$ & $0.17$ \\
CDTG$-10-$GI & $4$ & ($-21.2$,$18.5$,$-97.6$) & ($828.2$,$121.9$,$233.3$) & $-1.597$ & $0.94$ \\
 & & ($178.8$,$5.7$,$84.8$) & ($62.9$,$32.2$,$24.8$) & $0.027$ & $0.01$ \\
CDTG$-11-$GI & $4$ & ($-4.1$,$245.7$,$13.0$) & ($57.9$,$2028.9$,$42.9$) & $-1.496$ & $0.20$ \\
 & & ($55.3$,$18.4$,$29.1$) & ($6.7$,$76.8$,$31.0$) & $0.012$ & $0.02$ \\
CDTG$-12-$GI & $4$ & ($-100.5$,$-21.6$,$81.1$) & ($1195.5$,$-244.1$,$683.1$) & $-1.338$ & $0.89$ \\
 & & ($128.5$,$0.7$,$137.7$) & ($94.2$,$170.4$,$72.5$) & $0.039$ & $0.04$ \\
CDTG$-13-$GI & $4$ & ($-12.1$,$13.1$,$-67.9$) & ($491.5$,$96.7$,$232.5$) & $-1.789$ & $0.92$ \\
 & & ($52.3$,$12.6$,$38.1$) & ($23.2$,$93.2$,$23.1$) & $0.024$ & $0.05$ \\
CDTG$-14-$GI & $3$ & ($-51.7$,$-27.2$,$-1.2$) & ($1234.7$,$-322.1$,$97.5$) & $-1.449$ & $0.92$ \\
 & & ($164.9$,$6.9$,$32.9$) & ($85.5$,$70.6$,$24.0$) & $0.038$ & $0.02$ \\
CDTG$-15-$GI & $3$ & ($-8.3$,$63.5$,$-38.5$) & ($306.6$,$405.4$,$90.8$) & $-1.881$ & $0.71$ \\
 & & ($60.7$,$6.3$,$48.7$) & ($22.9$,$39.8$,$30.8$) & $0.020$ & $0.05$ \\
CDTG$-16-$GI & $3$ & ($87.5$,$32.5$,$-11.8$) & ($614.9$,$299.5$,$204.5$) & $-1.669$ & $0.86$ \\
 & & ($26.3$,$14.0$,$84.7$) & ($18.2$,$155.8$,$13.2$) & $0.037$ & $0.07$ \\
\enddata
\end{deluxetable*}

%% file: Tables/g1_cluster_mean_table.tex
\clearpage
\startlongtable
\begin{deluxetable*}{l  r  c  c  c  c  c  c  c}
\tablecaption{Group I CDTG Abundance Means, Dispersions and Inter-Quartile Ranges (IQR) \label{tab:g1_cluster_mean}}
\tablehead{\colhead{Cluster} & \colhead{$N$ Stars} & \colhead{[Fe/H]} & \colhead{[C/Fe]$_{c}$} & \colhead{[Mg/Fe]} & \colhead{[Sr/Fe]} & \colhead{[Y/Fe]} & \colhead{[Ba/Fe]} & \colhead{[Eu/Fe]}}
\startdata
CDTG$-1-$GI & $11$ & $-2.37\pm0.15$ &$+1.77\pm0.64$ &$+0.33\pm0.18$ &$+1.05\pm0.19$ &$+1.02\pm \dots$ &$+1.44\pm0.71$ &$+0.96\pm0.64$\\
CDTG$-2-$GI & $8$ & $-2.16\pm0.42$ &$+1.63\pm0.72$ &$+0.39\pm0.25$ &$+0.79\pm0.10$ &$+1.54\pm \dots$ &$+1.35\pm0.81$ &$+1.29\pm0.84$\\
CDTG$-3-$GI & $7$ & $-2.29\pm0.68$ &$+1.88\pm0.79$ &$+0.31\pm \dots$ &$+0.58\pm0.39$ &$\dots$ &$+2.13\pm0.68$ &$+1.44\pm0.71$\\
CDTG$-4-$GI & $6$ & $-2.39\pm0.54$ &$+2.04\pm0.51$ &$+0.29\pm0.16$ &$+0.71\pm0.07$ &$\dots$ &$+1.45\pm0.31$ &$+1.30\pm0.36$\\
CDTG$-5-$GI & $6$ & $-1.86\pm0.40$ &$+0.97\pm0.74$ &$+0.69\pm \dots$ &$+1.90\pm \dots$ &$+1.66\pm \dots$ &$+1.74\pm0.90$ &$+1.34\pm0.20$\\
CDTG$-6-$GI & $6$ & $-2.10\pm0.42$ &$+1.50\pm0.48$ &$+0.38\pm0.17$ &$+0.30\pm0.40$ &$\dots$ &$+1.13\pm0.94$ &$+0.94\pm0.72$\\
CDTG$-7-$GI & $5$ & $-2.96\pm0.18$ &$+2.20\pm0.65$ &$+0.40\pm0.13$ &$+0.22\pm0.21$ &$\dots$ &$+1.70\pm0.19$ &$+0.92\pm0.45$\\
CDTG$-8-$GI & $5$ & $-2.26\pm0.25$ &$+2.07\pm0.46$ &$+0.28\pm0.24$ &$+0.67\pm0.29$ &$\dots$ &$+1.75\pm0.73$ &$+1.31\pm0.74$\\
CDTG$-9-$GI & $5$ & $-2.12\pm0.14$ &$+1.37\pm0.31$ &$+0.19\pm \dots$ &$+0.92\pm0.60$ &$\dots$ &$+0.44\pm0.77$ &$+0.72\pm0.20$\\
CDTG$-10-$GI & $4$ & $-2.49\pm0.56$ &$+2.02\pm0.40$ &$+0.34\pm \dots$ &$+0.68\pm \dots$ &$\dots$ &$+1.47\pm1.52$ &$+1.56\pm0.30$\\
CDTG$-11-$GI & $4$ & $-1.58\pm0.47$ &$+1.70\pm0.68$ &$+0.80\pm \dots$ &$+1.50\pm0.79$ &$\dots$ &$+1.91\pm0.29$ &$+1.11\pm \dots$\\
CDTG$-12-$GI & $4$ & $-2.35\pm0.38$ &$+1.61\pm0.57$ &$+0.63\pm0.34$ &$+1.72\pm \dots$ &$\dots$ &$+1.63\pm0.38$ &$+0.68\pm \dots$\\
CDTG$-13-$GI & $4$ & $-2.24\pm0.55$ &$+2.09\pm0.72$ &$+0.64\pm \dots$ &$+0.48\pm0.30$ &$\dots$ &$+1.18\pm0.83$ &$+1.37\pm0.38$\\
CDTG$-14-$GI & $3$ & $-2.47\pm0.62$ &$+2.32\pm0.52$ &$+0.36\pm \dots$ &$+0.27\pm \dots$ &$\dots$ &$+1.83\pm \dots$ &$\dots$\\
CDTG$-15-$GI & $3$ & $-2.44\pm0.37$ &$+2.18\pm0.61$ &$+0.63\pm \dots$ &$+0.92\pm0.70$ &$\dots$ &$+0.79\pm0.98$ &$+0.68\pm0.22$\\
CDTG$-16-$GI & $3$ & $-2.50\pm0.22$ &$+2.11\pm0.74$ &$+0.37\pm \dots$ &$+0.79\pm \dots$ &$\dots$ &$+1.82\pm0.73$ &$+0.34\pm \dots$\\
\cline{0-8}
Biweight (CDTG mean): & & $-2.31\pm0.07$ & $+1.90\pm0.09$ & $+0.38\pm0.05$ & $+0.72\pm0.12$ & $+1.54\pm0.18$ & $+1.56\pm0.10$ & $+1.09\pm0.09$\\
Biweight (CDTG std): & & $+0.40\pm0.04$ & $+0.61\pm0.03$ & $\dots$ & $\dots$ & $\dots$ & $\dots$ & $\dots$\\
IQR (CDTG mean): & & $\boldsymbol{0.30}$ & $\boldsymbol{0.47}$ & $\boldsymbol{0.31}$ & $\boldsymbol{0.40}$ & $\boldsymbol{0.32}$ & $\boldsymbol{0.46}$ & $\boldsymbol{0.51}$\\
IQR (CDTG std): & & $0.30$ & $0.22$ & $\dots$ & $\dots$ & $\dots$ & $\dots$ & $\dots$\\
\cline{0-8}
Biweight (Final): & & $-2.57\pm0.59$ & $+1.13\pm0.63$ & $+0.43\pm0.21$ & $+0.17\pm0.71$ & $-0.06\pm0.47$ & $+0.17\pm1.13$ & $+0.51\pm0.65$\\
IQR (Final): & & $\boldsymbol{0.80}$ & $\boldsymbol{0.97}$ & $\boldsymbol{0.26}$ & $\boldsymbol{0.86}$ & $\boldsymbol{0.47}$ & $\boldsymbol{1.62}$ & $\boldsymbol{0.80}$\\
\enddata
\tablecomments{The first section of the table lists the biweight location and scale of the abundances for each of the CDTGs. The second section of the table lists the mean and the standard error of the mean (using biweight estimates) for both the location and scale of the abundances of the CDTGs, along with the IQR of the abundances of the CDTGs. The third section of the table lists the biweight location and scale of the Final Sample for each of the abundances, along with the IQR for each of the abundances in the Final Sample.}\end{deluxetable*}

%% file: Tables/group2_cluster_stellar_results_stub_table.tex
\begin{deluxetable*}{l  c  c  c  c  c  c  c  c}
\tablecaption{Group II CDTGs Identified by \texttt{HDBSCAN} \label{tab:g2_cluster_results_stub}}
\tablehead{\colhead{NAME} & \colhead{[Fe/H]} & \colhead{[C/Fe]$_{c}$} & \colhead{[Mg/Fe]} & \colhead{[Sr/Fe]} & \colhead{[Y/Fe]} & \colhead{[Ba/Fe]} & \colhead{[Eu/Fe]}}
\startdata
\multicolumn{8}{c}{CDTG$-1-$GII} \\
\multicolumn{8}{c}{Structure: Unassigned Structure} \\
\multicolumn{8}{c}{Group Assoc: C-3: \citet{Malhan_2022}} \\
\multicolumn{8}{c}{Group Assoc: NGC7089: \citet{Malhan_2022}} \\
\multicolumn{8}{c}{Group Assoc: DTG-9: \citet{Shank_2022a}} \\
\multicolumn{8}{c}{Group Assoc: DTG-110: \citet{Shank_2022b}} \\
\multicolumn{8}{c}{Group Assoc: DTG-117: \citet{Shank_2022b}} \\
\multicolumn{8}{c}{Stellar Assoc: HE 0420+0123a (DTC-3: \citealt{Hattori_2022})} \\
\multicolumn{8}{c}{Stellar Assoc: J042314.50+013048.0 (DTG-99: \citealt{Shank_2022b})} \\
\multicolumn{8}{c}{Stellar Assoc: J141640.80-242200.0 (DTG-43: \citealt{Shank_2022b})} \\
\multicolumn{8}{c}{Stellar Assoc: J220216.40-053648.0 (DTG-110: \citealt{Shank_2022b}) } \\
\multicolumn{8}{c}{Globular Assoc: No Globular Associations} \\
\multicolumn{8}{c}{Dwarf Galaxy Assoc: No Dwarf Galaxy Associations} \\
RAVE J030639.1$-$692040 & $-2.79$ & $+0.79$ & $\dots$ & $-0.02$ & $\dots$ & $+0.20$ & $-0.12$ \\
HE 0420$+$0123a & $-3.03$ & $+0.78$ & $+0.44$ & $+0.11$ & $\dots$ & $+0.08$ & $+0.79$ \\
HE 1305$-$0331 & $-3.26$ & $+1.09$ & $\dots$ & $-0.07$ & $\dots$ & $\dots$ & $\dots$ \\
2MASS J14164084$-$2422000 & $-2.57$ & $+0.87$ & $+0.37$ & $+0.02$ & $\dots$ & $-0.31$ & $+0.14$ \\
SDSS J152953.94$+$080448.1 & $-3.18$ & $+0.81$ & $+0.22$ & $\dots$ & $\dots$ & $-1.01$ & $\dots$ \\
2MASS J19504989$-$3321107 & $-2.61$ & $+0.70$ & $-0.10$ & $-0.35$ & $\dots$ & $-0.65$ & $0.00$ \\
BPS CS 22873$-$0128 & $-3.24$ & $+0.71$ & $+0.59$ & $-0.50$ & $\dots$ & $-1.45$ & $-0.34$ \\
BPS CS 22956$-$0050 & $-3.57$ & $+0.89$ & $+0.57$ & $-0.54$ & $\dots$ & $-1.00$ & $-0.19$ \\
2MASS J22021636$-$0536483 & $-2.75$ & $+0.75$ & $\dots$ & $+0.15$ & $\dots$ & $-0.75$ & $-0.12$ \\
BPS CS 22886$-$0043 & $-2.40$ & $+0.71$ & $+0.55$ & $+1.01$ & $\dots$ & $+0.54$ & $+1.04$ \\
$\boldsymbol{\mu \pm \sigma ([X/Y])}$ & $\boldsymbol{-2.93\pm0.37}$ & $\boldsymbol{+0.79\pm0.10}$ & $\boldsymbol{+0.43\pm0.22}$ & $\boldsymbol{-0.10\pm0.39}$ & $\boldsymbol{\dots}$ & $\boldsymbol{-0.56\pm0.69}$ & $\boldsymbol{-0.09\pm0.37}$\\
\enddata
\tablecomments{$\boldsymbol{\mu}$ and $\boldsymbol{\sigma}$ represent the biweight estimates of the location and scale for the abundances in the CDTG.}\tablecomments{This table is a stub; the full table is available in the electronic edition.}\end{deluxetable*}

%% file: Tables/g2_cluster_orbital_table.tex
\clearpage
\startlongtable
\begin{deluxetable*}{l  r  c  c  r  r}
\tablecaption{Group II Cluster Dynamical Parameters Determined by \texttt{AGAMA} \label{tab:g2_cluster_orbit}}
\tablehead{\colhead{Cluster} & \colhead{$N$ Stars} & \colhead{($\langle$v$_{\text{r}}\rangle$,$\langle$v$_{\phi}\rangle$,$\langle$v$_{\text{z}}\rangle$)} & \colhead{($\langle$J$_{\text{r}}\rangle$,$\langle$J$_{\phi}\rangle$,$\langle$J$_{\text{z}}\rangle$)} & \colhead{$\langle$E$\rangle$} & \colhead{$\langle$ecc$\rangle$}\\
\colhead{} & \colhead{} & \colhead{($\sigma_{\langle\text{v}_{\text{r}}\rangle}$,$\sigma_{\langle\text{v}_{\phi}\rangle}$,$\sigma_{\langle\text{v}_{\text{z}}\rangle}$)} & \colhead{($\sigma_{\langle\text{J}_{\text{r}}\rangle}$,$\sigma_{\langle\text{J}_{\phi}\rangle}$,$\sigma_{\langle\text{J}_{\text{z}}\rangle}$)} & \colhead{$\sigma_{\langle\text{E}\rangle}$} & \colhead{$\sigma_{\langle\text{ecc}\rangle}$}\\
\colhead{} & \colhead{} & \colhead{(km s$^{-1}$)} & \colhead{(kpc km s$^{-1}$)} & \colhead{(10$^{5}$ km$^{2}$ s$^{-2}$)} & \colhead{}}
\startdata
CDTG$-1-$GII & $10$ & ($-17.0$,$88.7$,$1.6$) & ($230.6$,$587.7$,$262.0$) & $-1.754$ & $0.57$ \\
 & & ($86.2$,$26.2$,$82.4$) & ($95.1$,$66.3$,$97.0$) & $0.069$ & $0.09$ \\
CDTG$-2-$GII & $8$ & ($-60.2$,$-49.4$,$166.6$) & ($309.5$,$-323.5$,$641.9$) & $-1.691$ & $0.61$ \\
 & & ($84.6$,$36.8$,$28.4$) & ($69.3$,$242.8$,$37.9$) & $0.093$ & $0.10$ \\
CDTG$-3-$GII & $4$ & ($-33.5$,$64.4$,$6.4$) & ($406.1$,$542.0$,$63.7$) & $-1.771$ & $0.72$ \\
 & & ($14.6$,$10.3$,$60.8$) & ($23.0$,$85.1$,$42.9$) & $0.020$ & $0.04$ \\
CDTG$-4-$GII & $4$ & ($-20.8$,$44.9$,$-10.9$) & ($613.8$,$354.4$,$66.7$) & $-1.728$ & $0.83$ \\
 & & ($135.8$,$7.9$,$14.1$) & ($71.6$,$41.1$,$68.7$) & $0.021$ & $0.01$ \\
CDTG$-5-$GII & $4$ & ($0.4$,$169.3$,$-45.9$) & ($76.3$,$1314.0$,$80.8$) & $-1.649$ & $0.25$ \\
 & & ($1.3$,$19.9$,$13.1$) & ($59.8$,$65.5$,$33.5$) & $0.009$ & $0.11$ \\
CDTG$-6-$GII & $3$ & ($-25.3$,$19.5$,$-20.7$) & ($569.7$,$146.3$,$46.6$) & $-1.829$ & $0.90$ \\
 & & ($59.1$,$7.8$,$48.5$) & ($40.9$,$58.8$,$19.7$) & $0.012$ & $0.06$ \\
CDTG$-7-$GII & $3$ & ($5.0$,$45.8$,$14.2$) & ($775.0$,$425.4$,$102.9$) & $-1.600$ & $0.83$ \\
 & & ($151.5$,$7.1$,$52.6$) & ($46.8$,$29.4$,$42.3$) & $0.044$ & $0.02$ \\
CDTG$-8-$GII & $3$ & ($17.6$,$125.8$,$20.0$) & ($343.9$,$1182.7$,$66.3$) & $-1.595$ & $0.53$ \\
 & & ($97.2$,$4.2$,$39.8$) & ($40.6$,$114.6$,$47.9$) & $0.028$ & $0.03$ \\
CDTG$-9-$GII & $3$ & ($-82.1$,$124.4$,$-83.6$) & ($396.7$,$610.2$,$902.0$) & $-1.475$ & $0.55$ \\
 & & ($85.6$,$40.8$,$109.0$) & ($94.2$,$185.1$,$52.5$) & $0.035$ & $0.05$ \\
\enddata
\end{deluxetable*}

%% file: Tables/g2_cluster_mean_table.tex
\clearpage
\startlongtable
\begin{deluxetable*}{l  r  c  c  c  c  c  c  c}
\tablecaption{Group II CDTG Abundance Means, Dispersions and Inter-Quartile Ranges (IQR) \label{tab:g2_cluster_mean}}
\tablehead{\colhead{Cluster} & \colhead{$N$ Stars} & \colhead{[Fe/H]} & \colhead{[C/Fe]$_{c}$} & \colhead{[Mg/Fe]} & \colhead{[Sr/Fe]} & \colhead{[Y/Fe]} & \colhead{[Ba/Fe]} & \colhead{[Eu/Fe]}}
\startdata
CDTG$-1-$GII & $10$ & $-2.93\pm0.37$ &$+0.79\pm0.10$ &$+0.43\pm0.22$ &$-0.10\pm0.39$ &$\dots$ &$-0.56\pm0.69$ &$-0.09\pm0.37$\\
CDTG$-2-$GII & $8$ & $-3.23\pm0.41$ &$+0.85\pm0.13$ &$+0.50\pm0.10$ &$+0.05\pm0.51$ &$+0.20\pm \dots$ &$-0.76\pm0.61$ &$+0.28\pm0.26$\\
CDTG$-3-$GII & $4$ & $-2.96\pm0.21$ &$+0.91\pm0.19$ &$+0.41\pm0.02$ &$-0.13\pm0.42$ &$\dots$ &$-0.56\pm0.30$ &$+1.10\pm0.37$\\
CDTG$-4-$GII & $4$ & $-3.17\pm0.45$ &$+0.96\pm0.16$ &$+0.29\pm0.07$ &$-0.22\pm0.26$ &$\dots$ &$-1.32\pm0.27$ &$+0.41\pm0.52$\\
CDTG$-5-$GII & $4$ & $-3.03\pm0.19$ &$+0.91\pm0.22$ &$+0.32\pm \dots$ &$-0.63\pm0.60$ &$\dots$ &$-0.52\pm0.26$ &$+0.29\pm \dots$\\
CDTG$-6-$GII & $3$ & $-2.85\pm0.22$ &$+0.74\pm0.02$ &$+0.44\pm \dots$ &$-0.29\pm \dots$ &$\dots$ &$-0.49\pm \dots$ &$+0.19\pm0.86$\\
CDTG$-7-$GII & $3$ & $-3.08\pm0.45$ &$+0.84\pm0.07$ &$+0.39\pm \dots$ &$-0.23\pm \dots$ &$\dots$ &$-0.13\pm \dots$ &$+1.79\pm \dots$\\
CDTG$-8-$GII & $3$ & $-3.35\pm0.46$ &$+0.94\pm0.17$ &$+0.58\pm \dots$ &$-1.19\pm \dots$ &$\dots$ &$-0.58\pm0.76$ &$+0.17\pm \dots$\\
CDTG$-9-$GII & $3$ & $-3.15\pm0.26$ &$+1.12\pm0.30$ &$+0.64\pm0.17$ &$-0.41\pm0.52$ &$\dots$ &$-0.45\pm0.77$ &$+0.20\pm0.48$\\
\cline{0-8}
Biweight (CDTG mean): & & $-3.08\pm0.05$ & $+0.89\pm0.04$ & $+0.44\pm0.04$ & $-0.23\pm0.08$ & $+0.20\pm0.00$ & $-0.55\pm0.04$ & $+0.23\pm0.07$\\
Biweight (CDTG std): & & $+0.34\pm0.04$ & $+0.15\pm0.03$ & $\dots$ & $\dots$ & $\dots$ & $\dots$ & $\dots$\\
IQR (CDTG mean): & & $\boldsymbol{0.21}$ & $\boldsymbol{0.10}$ & $\boldsymbol{0.11}$ & $\boldsymbol{0.28}$ & $\boldsymbol{0.00}$ & $\boldsymbol{0.09}$ & $\boldsymbol{0.22}$\\
IQR (CDTG std): & & $0.23$ & $0.09$ & $\dots$ & $\dots$ & $\dots$ & $\dots$ & $\dots$\\
\cline{0-8}
Biweight (Final): & & $-2.57\pm0.59$ & $+1.13\pm0.63$ & $+0.43\pm0.21$ & $+0.17\pm0.71$ & $-0.06\pm0.47$ & $+0.17\pm1.13$ & $+0.51\pm0.65$\\
IQR (Final): & & $\boldsymbol{0.80}$ & $\boldsymbol{0.97}$ & $\boldsymbol{0.26}$ & $\boldsymbol{0.86}$ & $\boldsymbol{0.47}$ & $\boldsymbol{1.62}$ & $\boldsymbol{0.80}$\\
\enddata
\tablecomments{The first section of the table lists the biweight location and scale of the abundances for each of the CDTGs. The second section of the table lists the mean and the standard error of the mean (using biweight estimates) for both the location and scale of the abundances of the CDTGs, along with the IQR of the abundances of the CDTGs. The third section of the table lists the biweight location and scale of the Final Sample for each of the abundances, along with the IQR for each of the abundances in the Final Sample.}\end{deluxetable*}